\documentclass{emulateapj}

\usepackage{graphicx}
\usepackage{txfonts}
\usepackage{amssymb}
\usepackage{enumerate}
\usepackage[usenames,dvipsnames]{color}           
\newcommand{\BE}{\begin{equation}}
\newcommand{\EE}{\end{equation}}
\newcommand{\BA}{\begin{eqnarray}}
\newcommand{\EA}{\end{eqnarray}}

\renewcommand{\vec}[1]{{\mathbf #1}}

\newcommand{\avec}{ \vec A}

\newcommand{\bb}{\vec B}

\newcommand{\jj}{ \vec j}

\newcommand{\xx}{ \vec x}

\newcommand{\eq}[1]{Equation~(\ref{eq:#1})} 
\newcommand{\eqs}[2]{Equations~(\ref{eq:#1}) and (\ref{eq:#2})} 
\newcommand{\eqss}[2]{Equations~(\ref{eq:#1}) -- (\ref{eq:#2})} 
\newcommand{\sect}[1]{Section~\ref{sec:#1}}

\newcommand{\tab}[1]{Table~\ref{tab:#1}}

\newcommand{\fig}[1]{Figure~\ref{fig:#1}}

\newcommand{\cf}{\textit{cf.} }
\newcommand{\eg}{\textit{e.g.}, }
\newcommand{\ie}{\textit{i.e.}, }

\shorttitle{Data-Optimized Coronal Field Model}
\shortauthors{Dalmasse et al.}

\begin{document} 

   \title{Data-Optimized Coronal Field Model:\\
          I. Proof of concept}
   
   \author{
	K. Dalmasse$^{1}$,
	A. Savcheva$^{2,3}$,
	S.~E. Gibson$^{4}$,
	Y. Fan$^{4}$,
	D.~W. Nychka$^{5}$,
	N. Flyer$^{4}$,
	N. Mathews$^{6}$,
	E.~E. DeLuca$^{2}$
	}
		
   \affil{$1$ IRAP, Universit\'e de Toulouse, CNRS, CNES, UPS, 
                 31028 Toulouse, France
                 }
	\email{kevin.dalmasse@irap.omp.eu}
   \affil{$2$ Harvard-Smithsonian Center for Astrophysics, 
   				 60 Garden Street, Cambridge, MA 02138, USA
                 }
   \affil{$3$ Institute of Astronomy and National Astronomical Observatory, 
   				 Bulgarian Academy of Sciences, 72 Tsarigradsko Chaussee Blvd., 1784 Sofia, Bulgaria
                 }                
   \affil{$4$ National Center for Atmospheric Research, 
                 P.O. Box 3000, Boulder, CO 80307-3000, USA
                 }
   \affil{$5$ Department of Applied Mathematics and Statistics, 
                 Colorado School of Mines, 1500 Illinois St., Golden, CO 80401, USA
                 }
   \affil{$6$ Department of Applied Mathematics, 
   				University of Colorado at Boulder, 526 UCB, Boulder, CO 80309-0526, USA
                 }

 
   \begin{abstract}
Deriving the strength and direction of the three-dimensional (3D) magnetic field in the solar 
atmosphere is fundamental for understanding its dynamics. Volume information on the magnetic 
field mostly relies on coupling 3D reconstruction 
methods with photospheric and/or chromospheric surface vector magnetic fields. Infrared 
coronal polarimetry could provide additional information to better constrain magnetic 
field reconstructions. However, combining such data with reconstruction methods is 
challenging, \eg because of the optical-thinness of the solar corona and the lack and 
limitations of stereoscopic polarimetry. To address these issues, we introduce 
the Data-Optimized Coronal Field Model (DOCFM) framework, a model-data fitting approach 
that combines a parametrized 3D generative model, \eg a magnetic field extrapolation or 
a magnetohydrodynamic model, with forward modeling of coronal data. We test it 
with a parametrized flux rope insertion method and infrared coronal polarimetry 
where synthetic observations are created from a known ``ground truth'' physical state. 
We show that this framework allows us to 
accurately retrieve the ground truth 3D magnetic field of a set of force-free field 
solutions from the flux rope insertion method. 
In observational studies, the DOCFM will provide a means to force 
the solutions derived with different reconstruction methods to satisfy additional, common, 
coronal constraints. The DOCFM framework therefore opens new perspectives for the exploitation 
of coronal polarimetry in magnetic field reconstructions and for developing new techniques 
to more reliably infer the 3D magnetic fields that trigger solar flares and coronal mass ejections.
   \end{abstract}   

   \keywords{polarization -- Sun: corona -- Sun: magnetic fields
               }

%

\section{Introduction} \label{sec:S-Introduction}

%
%
Solar flares and coronal mass ejections (CMEs) are driven by the evolution 
of current-carrying magnetic fields in the solar corona 
\citep[\eg][]{Forbes00,Priest03,Schrijver05,Shibata11,Aulanier12}. 
Deriving the three-dimensional (3D) properties of such non-potential magnetic 
fields is critical for identifying the mechanism(s) driving flares and CMEs, 
as well as for understanding and predicting their evolution 
\citep[\eg][]{Bateman78,Hood81,Antiochos99,Kusano12,Pariat17}. Hence, measuring the strength 
and direction of the 3D magnetic field in the solar coronal volume is fundamental.

%
%
Magnetic field information in the solar corona 
{is}
mostly derived from the inversion 
of off-limb polarization measurements associated with the Zeeman and Hanle effects 
\citep[\eg][]{Harvey69,Casini99,Lin04,Centeno10}. The Zeeman effect creates 
a frequency-modulated polarization signal sensitive to both the strength and direction 
of the magnetic field. Due to the large Doppler widths of coronal emission lines 
and the wavelength squared scaling, the Zeeman effect in the corona is better observed 
with infrared (IR) spectral lines \citep[\eg][]{Judge98,Penn14}. However, the coronal 
magnetic field is weak, with typical values of 1 to 10 Gauss, except right above solar 
active regions where it can reach up to a few 100 Gauss \citep[\eg][]{Kuhn96,Lin00}. 
The corresponding fraction of circular polarization in IR lines, such as the Fe XIII 
lines, is thus only expected to be of the order of $10^{-4}$ 
\citep[equivalent to a 1 Gauss magnetic field strength; \eg][]{Querfeld82,Plowman14}. 
Accurately measuring the Zeeman-induced polarization signal in the corona is therefore 
a challenging task that will require large aperture telescopes, such as the Large 
Coronagraph (1.5 meter) on the COronal Solar Magnetism Observatory \citep[COSMO;][]{Tomczyk16} 
or the 4-meter Daniel K. Inouye Solar Telescope \citep[DKIST; see][and references therein]{Keil11}.

%
%
The Hanle effect is the second main mechanism exploited for diagnosing the solar 
coronal magnetic field. This process modifies 
{the polarization}
of spectral 
lines in the presence of a magnetic field \citep[\eg][]{Hanle24,SahalBrechot77,Bommier82,Arnaud87}. 
As opposed to the Zeeman effect, the Hanle effect is a depolarization mechanism. 
It therefore requires the prior existence of polarization by means of other physical 
processes such as, \eg radiation scattering \citep[\eg][]{Charvin65}. 
Sensitivity of the Hanle effect to the magnetic field can range from a few milli-Gauss 
to several hundred Gauss depending on the choice of spectral line and the strength 
and direction of the magnetic field \citep[\eg][]{Bommier82,Raouafi16}. The Hanle 
effect is hence a powerful tool for probing the coronal magnetic field, as confirmed 
by theoretical \citep[\eg][]{Judge06,Rachmeler13,Rachmeler14,Dalmasse16} and 
observational \citep[\eg][]{BakSteslicka13,Morton16,Gibson17,KarnaSub} studies 
with, \eg off-limb coronal polarimetry in the IR Fe XIII lines. Note, though, that 
routine measurements of coronal polarization are currently limited to the IR Fe XIII 
lines with the Coronal Multi-channel Polarimeter \citep[CoMP;][]{Tomczyk08} for which 
the Hanle effect operates in the saturated regime \citep[\eg][]{Casini99}. In practice, 
it means that the measured linear polarization is sensitive to the magnetic field 
direction but not its strength. Coronal polarimetry with other spectral lines, such 
as the IR He I 10830 \AA\ or the UV H I Ly $\alpha$ lines, will be necessary to further 
obtain Hanle diagnostics sensitive to the coronal magnetic field strength \citep[\eg][]{Raouafi16}.

%
%
While the Zeeman and Hanle effects offer powerful diagnostics of the coronal magnetic 
field, determining the actual 3D coronal magnetic field from coronal polarimetry remains 
a true challenge. In addition to the previously discussed limitations, the solar corona 
is optically thin at most wavelengths. It follows that the measured polarization signal 
is the integration of all the plasma emission along the line of sight (LOS). Consequently, 
it is in general not possible to invert the polarization maps into 2D maps of the magnetic 
field. Furthermore, it is difficult to extract individual magnetic field data at specific 
positions along the LOS, even with stereoscopic measurements whether used on their own or 
combined with 3D magnetic field extrapolation methods \citep[\eg][]{Kramar13,Kramar16}. 
In particular, one of the main challenges with stereoscopic polarimetry relies on the limited 
amount of information that can be retrieved from the data, either due to a limited range 
in magnetic strength sensitivity at a given wavelength, or the lack of it for, e.g., 
the linear polarization signal measured for the Fe XIII lines. The latter is only sensitive 
to the magnetic field direction and further possesses both a $90^{\circ}$ and $180^{\circ}$ 
ambiguity \citep[saturated regime of the Hanle effect;][]{Judge07,Plowman14}.

  \begin{figure*}
   \centerline{\includegraphics[width=0.95\textwidth,clip=]{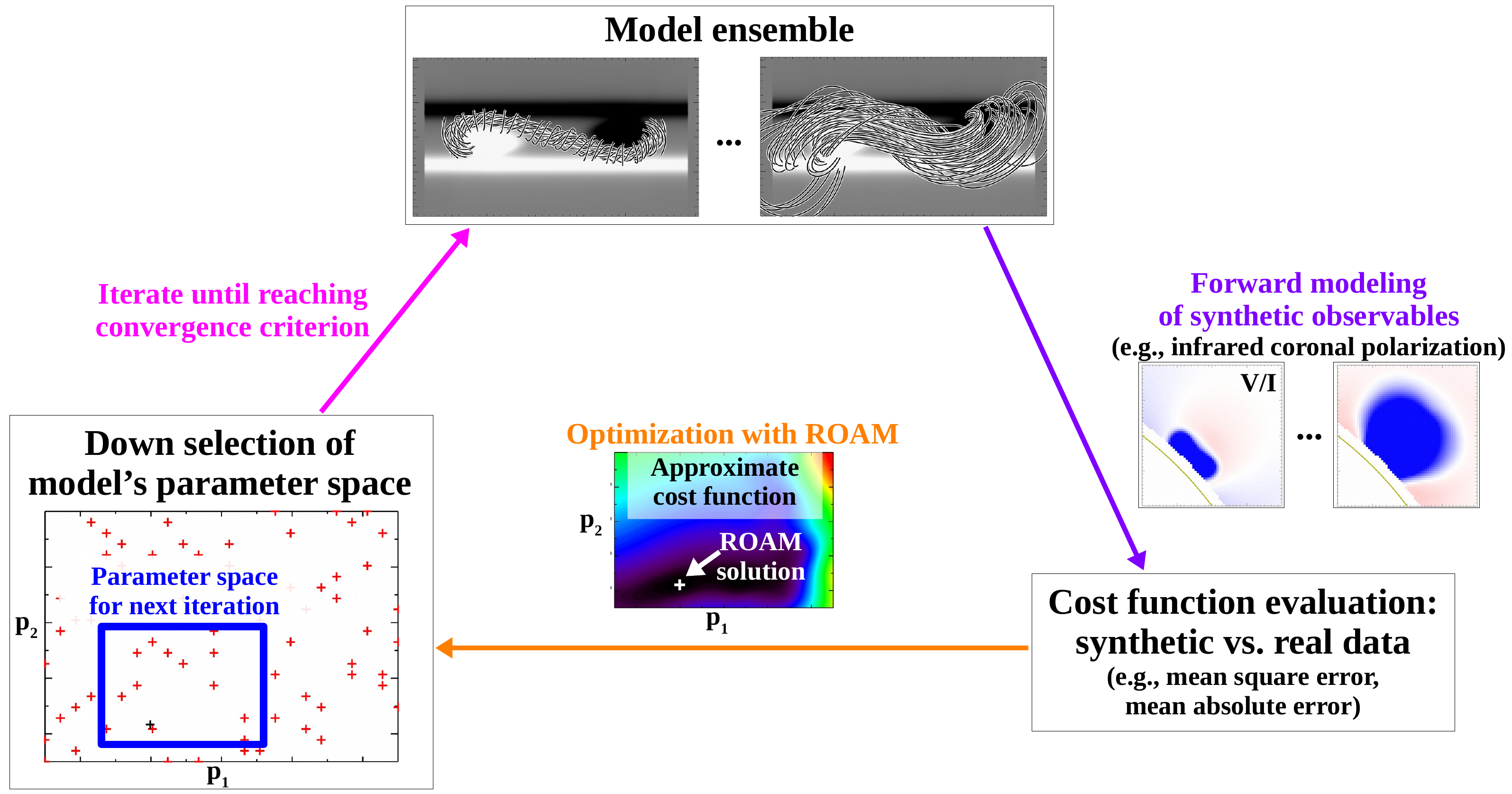}
              }
   \caption{General sketch of the DOCFM framework.}
              \label{fig:Fig-DOCFM-diagram}
   \end{figure*} 

%
%
Volume information on the vector magnetic field in the solar corona thus mostly relies 
on the approximate 3D solution obtained by coupling surface magnetic field maps (so-called 
vector magnetograms), derived from photospheric and/or chromospheric polarimetry, with 3D 
magnetic field reconstruction methods. 3D techniques for reconstructing the solar coronal 
magnetic field from surface measurements are either of the nonlinear force-free field 
\citep[NLFFF; \eg][]{vanBallegooijen04,Wheatland07,Wiegelmann10,Valori10,Contopoulos11,Malanushenko12,Amari13,Yeates14} 
or the magnetohydrodynamics \citep[MHD; \eg][]{Mikic99,Inoue11,Feng12,Zhu13} type.

These reconstruction methods differ by the equations solved, the implemented algorithms, 
and their treatment of the vector magnetograms as boundary conditions (\eg full vs. 
LOS vector, pre-processing to create more force-free boundary conditions). 
As a consequence, the 3D, current-carrying, magnetic field solution and its properties 
can strongly vary from one method to the other. For instance, \cite{DeRosa09}, 
\cite{DeRosa15} and \cite{Yeates18} reported variations between reconstruction methods 
that can reach up to 200$\%$ for the ratio of free to potential magnetic energy, 
as well as for relative magnetic helicity, which are 
key ingredients for producing solar flares and CMEs \citep[\eg][]{Low96,Forbes06,Tziotziou12,Zuccarello18}. 
Even the magnetic topology, which can also play a key role on the stability 
of the 3D magnetic field configuration \citep[\eg][]{Gorbachev89,Somov93,Savcheva16}, 
can be strongly affected by the choice of reconstruction method. This can make it 
difficult to determine the role of magnetic topology in the triggering of solar 
flares, in particular for events for which some reconstruction methods may produce 
a flux rope prior to the flare \citep[\eg][]{Amari18} when others only produce sheared 
arcades \citep[\eg][]{Jiang16}.

%
%
The present paper is the first in a series that investigates the possibility of 
improving the reliability of 3D magnetic field reconstructions by further exploiting 
coronal polarimetry. The methodology we propose and the tools we use are described 
in \sect{S-Method}. \sect{S-Test-Case} presents the test-case for which we test and 
prove the concept of our approach, using a known ``ground truth'' 3D magnetic 
field to create synthetic observations. 
The results of our analysis are reported in \sect{S-Results}. 
\sect{S-Discussion} discusses the applicability, limitations and perspectives 
of our approach. Our conclusions are summarized in \sect{S-Conclusions}.


\section{Method} \label{sec:S-Method}

\subsection{Summary of approach} \label{sec:S-DOCFM}

%
%
The general framework we propose to couple existing 3D magnetic field reconstructions 
with coronal polarimetric observations is the Data-Optimized Coronal Field Model (DOCFM). 
It is a model-data fitting approach of the 3D reconstruction of the coronal magnetic 
field. The DOCFM is built on the three following general bases:
\begin{enumerate}
	\item A generative 3D magnetic field model (\ie extrapolation/reconstruction) parametrized 
			through its electric currents. This can be done either at the photospheric boundary 
			(\eg by means of the transverse magnetic field or the force-free parameter) or in the volume 
			\citep[\eg for flux rope insertion methods; \eg][]{vanBallegooijen04,Titov14,Titov18}. 
			The generative model then creates the physical state of the corona (\eg magnetic field, 
			plasma pressure, density, temperature).
	\item Forward modeling of coronal polarimetry, in particular to address the fact that the solar 
			corona is optically thin and the lack and limitations of stereoscopic coronal polarimetry.
	\item Finding the set of parameters that minimize a cost function 
			\citep[or maximize a likelihood function as in][]{Dalmasse16}, here the mean squared 
			error between the polarization signal predicted for the magnetic field model 
			and the real polarization data.
\end{enumerate}
\fig{Fig-DOCFM-diagram} presents a general chart of the DOCFM approach.

%
%
Previous work has demonstrated the sensitivity of polarization signals in coronal 
cavities to the 3D magnetic field geometry \citep[\eg][]{Judge06,BakSteslicka13,Rachmeler13,Rachmeler14}. 
Our main focus is thus on exploiting coronal cavities and their IR polarimetric 
signatures as observed by CoMP to constrain the 3D magnetic field that precedes 
CMEs. The parametrized 3D magnetic field model we choose to work with is the flux 
rope insertion method of \cite{vanBallegooijen04} briefly presented in \sect{S-Generative-Model}. 
This method uses the LOS measurement of the photospheric magnetic field and an analytical 
model to produce a coronal flux rope. It is particularly useful for studying coronal cavities, 
which are likely associated with 3D magnetic flux ropes in weak field regions where the LOS magnetic 
field is still well-enough above the noise level while the transverse field is not. The flux rope 
insertion method is thus better suited for studying such structures than more traditional extrapolation 
techniques for which flux ropes arise from the electric currents associated with the photospheric 
transverse magnetic field measurements, and which would likely fail to retrieve the flux ropes 
when such measurements are too noisy. 
Our choice of synthetic coronal polarimetric data is IR polarimetry in the 10747 \AA\ 
Fe XIII-I line, synthesized with the codes of the FORWARD\footnote{http://www.hao.ucar.edu/FORWARD/} 
IDL package \citep[see \sect{S-FORWARD};][]{Gibson16}. For the minimization, we use 
an iterative implementation of the Radial-basis-functions Optimization Approximation 
Method \citep{Dalmasse16} described in \sect{S-IROAM}.


\subsection{Generative magnetic field model} \label{sec:S-Generative-Model}

%
%
In the the strong field regions of the solar corona, the plasma $\beta$, \ie the ratio 
of plasma pressure to magnetic pressure, is relatively low 
\citep[about $10^{-4}$ to $10^{-2}$; \eg][]{Gary01}. In this low-$\beta$ environment, 
all non-magnetic forces (\eg kinematic plasma flow pressure, gravity) are dominated 
by the magnetic ones and can thus be neglected. As a consequence, the Lorentz force 
vanishes (\ie the magnetic pressure force is compensated by the magnetic tension force) 
and the coronal magnetic field can be modeled as a force-free field, such that
\BE
	\label{eq:Eq-FFF}
	\nabla \times \bb (r) = \frac{4 \pi}{c} \jj (r) = \alpha (r) \bb (r)      \,,
\EE
where $c$ is the speed of light, $\bb (r)$ is the vector magnetic field, $\jj (r)$ is 
the electric current density and $\alpha (r)$ is the force-free parameter. $\alpha (r)=0$ 
refers to the potential field solution, while constant-$\alpha$ solutions are the so-called 
linear force-free fields \citep[\eg][]{Alissandrakis81}. In the most general case, 
$\alpha (r)$ is constant along individual magnetic field lines, but varies from one field 
line to the other, which corresponds to the NLFFF solutions \citep[see, \eg review by][]{Wiegelmann12R}.

%
%
Several NLFFF methods have been developed to solve for \eq{Eq-FFF} and extrapolate, 
or reconstruct, the 3D coronal magnetic field from 2D photospheric 
magnetic field measurements as a bottom boundary condition 
\citep[see, \eg][to cite a few]{Grad58,Amari06,Wheatland07,Valori10,Inoue12,Malanushenko12,Wiegelmann12,Titov18}. 
In this paper, we use the flux rope insertion method of \cite{vanBallegooijen04}. 
Such a choice is motivated by the fact that (1) our main focus is on applying 
the DOCFM approach to coronal cavities, which are density-depleted regions 
likely associated with a 3D magnetic flux rope \citep[\eg][]{BakSteslicka13,Gibson15}, 
and (2) the flux rope insertion method is already a parametrized 3D generative 
model. The flux rope possesses two parameters, which are the axial flux (\ie 
the magnetic flux along the flux rope axis), $\Phi$, and the poloidal flux 
per unit length (\ie the magnetic flux per unit length in the direction 
perpendicular to the flux rope axis), F.

%
%
To apply the flux rope insertion method \citep{vanBallegooijen04}, a potential field 
source surface (PFSS) extrapolation is first computed from LOS photospheric magnetograms. 
EUV data of, \eg solar filaments \citep[see, \eg][]{Su11,Savcheva12}, are then used to determine 
the photospheric feet and path of the flux rope. A field-free (\ie zero-magnetic field), 3D 
thin channel is created along the flux rope path in the potential field. The parametrized 
3D flux rope is then inserted into that field-free thin channel, thus producing a magnetic 
field configuration containing a flux rope, but which is out of equilibrium. The magnetic 
field is then driven towards a quasi-force-free state by means of magnetofrictional relaxation 
\citep[more details on the flux rope insertion method can be found in, \eg][]{vanBallegooijen04,Bobra08,Savcheva16}. 
The entire flux rope insertion procedure is performed in terms of modifying and evolving 
the vector potential, $\avec$ (defined by $\bb = \nabla \times \avec$), 
through the induction equation and using hyperdiffusion to smooth gradients 
in the force-free parameter \citep{Yang86,vanBallegooijen00}. The advantage of performing 
the relaxation with the vector potential is that it automatically ensures 
that the solenoidal condition for the magnetic field (\ie $\nabla \cdot \bb = 0$) is 
numerically satisfied.

For all flux rope insertions produced in this paper, the magnetofrictional relaxation 
is applied without any diffusion for the first 100 steps. Then, hyperdiffusion is used until 
$6 \times 10^4$ relaxation steps have been performed, at which point the relaxation is stopped 
and stable NLFFF models are obtained.


\subsection{Forward modeling of coronal polarimetry} \label{sec:S-FORWARD}

%
%
As mentioned \sect{S-Introduction}, the solar corona is optically thin and stereoscopic 
observations of the coronal polarization are currently not available. In addition, 
the linear polarization measured in the Fe XIII lines is sensitive to the direction 
of the coronal magnetic field but not its strength. Hence, off-limb coronal polarimetry 
as measured by CoMP provides a LOS-integrated signal that cannot be inverted into 
a 2D plane-of-sky (POS) magnetic field that could be directly plugged in the 3D magnetic 
field model. To couple such coronal polarimetric data with coronal magnetic field 
reconstruction models, we must instead rely on a model-data fitting approach. The latter 
requires a means to produce synthetic coronal observations from a given magnetic field 
model cube and observing position.

%
%
To produce synthetic coronal polarimetric data, we use the FORWARD IDL suite \citep{Gibson16}.
FORWARD is a package for multiwavelength coronal magnetometry that is integrated into 
the SolarSoft\footnote{http://www.lmsal.com/solarsoft/} \citep{Freeland98} IDL toolset. 
It is designed to create synthetic observables and compare them to coronal data 
\citep[a full description is provided in][]{Gibson16}. In particular, FORWARD 
employs the Coronal Line Emission (CLE) polarimetry code developed by \cite{Casini99} 
to synthesize full Stokes $(I, Q, U, V)$ line profiles for visible and IR forbidden 
lines including, but not limited to, the Fe XIII lines routinely observed by the CoMP 
and used in the analyses performed in this paper. Stokes $I$ corresponds to the integrated 
total line intensity. Stokes $(Q,U)$ are the two components of the linear polarization. 
And Stokes $V$ is the circular polarization.


\subsection{Optimization of flux-rope parameters} \label{sec:S-IROAM}

%
%
To find the set of parameters minimizing the mean squared error (MSE) between the predicted 
polarization signal for the magnetic field model and the real data, we use an iterative 
implementation of the Radial-basis-functions Optimization Approximation Method 
\citep[ROAM; \eg][]{Dalmasse16}. 
ROAM relies on evaluating a cost function\footnote{In essence, minimizing 
a cost function can be seen as maximizing a (log)-likelihood. ROAM can therefore be used 
equivalently for maximizing a log-likelihood as in \cite{Dalmasse16} or minimizing a cost 
function as in this paper.} (hence, the generative magnetic field model and synthetic 
observables) for a sparse sample of model-parameter values, approximating the sparse cost 
function sample with a series of radial basis functions (RBF) to obtain an analytical form 
for the cost function as a function of the model-parameter values, and computing 
an estimate of the best-fit parameters by minimizing the analytical form of the cost 
function \citep[the more detailed procedure is provided in Section 2 of][]{Dalmasse16}.

%
%
The generative magnetic field model we use to test the applicability and accuracy 
of the DOCFM approach only has two parameters (\cf \sect{S-Generative-Model}). 
ROAM was developed to be a fast and efficient optimization method for higher-dimensional 
optimization problems, \ie with a number of parameters at least equal to 3. Thus, in practice, 
there exists other optimization methods that would be faster and more efficient (\ie requiring 
{fewer} 
model evaluations) than ROAM for the optimization problem at hand. However, the goal 
of this paper is to show the applicability of the complete DOCFM framework, which we developed 
to be general enough to be used with generative magnetic field models having a large number 
of parameters and for which ROAM is better suited. The latter is the motivation behind the use 
of ROAM in this investigation.

%
%
{Let $p$ be the number of (model)-parameters, $\xx = (x_1, x_2, ...,x_p)$ a vector parameter 
in the $p$-dimensional parameter space, and $S = \{ \xx^{i=\{1,...,n \}} \}$ a random sample 
of $n$ independent vector parameters (\ie $x^{j \ne i}_{k} \ne x^{i}_{k}$ for all $k$).} As shown 
by the tests performed in \cite{Dalmasse16}, the estimate of best-fit parameters obtained 
with the ROAM can be sensitive to the choice of sample of vector parameters, 
{\ie to the choice of $S$.} 
On the other hand, the mean best-fit vector parameter obtained by averaging the best-fit 
vector parameters computed by applying ROAM with different samples, actually provides a good 
approximation of the true best-fit vector parameter. We use the latter 
result to build an iterative implementation of ROAM that gives an estimate of 
the best-fit vector parameters that is quasi-independent of the sample choice. 
Our iterative application of ROAM is based on adaptive refinement, such that 
refinement is performed in the region where the best-fit vector parameter is 
likely to be. The algorithm is as follows:
\begin{enumerate}
	\item Let $R$ refer to both the refinement level and iteration number, such that $R=0$ corresponds 
			to the un-refined starting grid.
	\item Choose boundaries, $((x^{\mathrm{min}}_1,x^{\mathrm{max}}_1), ..., (x^{\mathrm{min}}_p,x^{\mathrm{max}}_p)) = ((x^{\mathrm{min},R=0}_1,x^{\mathrm{max},R=0}_1), ..., (x^{\mathrm{min},R=0}_p,x^{\mathrm{max},R=0}_p))$, 
			to define the parameter space region where to search for the parameter values minimizing the MSE.
	\item Use {{\it latin hypercube sampling} \citep[LHS; \eg][]{McKay79,Iman81}} to create 3, randomly 
			chosen, sparse samples, $(S^{R}_1, S^{R}_2, S^{R}_3)$ of $n$ vector 
			parameters for the parameter space region of interest. The 3 samples are chosen 
			with the constraint that the 3$n$ vector parameters are all different.
	\item Combine the latin hypercube samples to produce the 7 samples, 
			$(S^{\prime}_1, S^{\prime}_2, S^{\prime}_3, S^{\prime}_4, S^{\prime}_5, S^{\prime}_6, S^{\prime}_7) = (S^{R}_1, S^{R}_2, S^{R}_3, S^{R}_1 + S^{R}_2, S^{R}_1 + S^{R}_3, S^{R}_2 + S^{R}_3, S^{R}_1 + S^{R}_2 + S^{R}_3)$.
	\item Apply ROAM to the 7 samples, $(S^{\prime}_{d=1,...,7})$, to obtain 7 estimates of best-fit vector 
			parameters, $(\xx^{\mathrm{bf},R} (S^{\prime}_1) , ..., \xx^{\mathrm{bf},R} (S^{\prime}_7))$, 
			at the $R$-th iteration.
	\item Compute the optimization solution vector defined as the mean best-fit solution, 
			$\mu^{R} = (\mu^{R}_{1}, \mu^{R}_{2}, ..., \mu^{R}_{p})$, 
			and its standard deviation vector, 
			$\sigma^{R} = (\sigma^{R}_{1}, \sigma^{R}_{2}, ..., \sigma^{R}_{p})$, 
			at the $R$-th iteration
			\BA
				\label{eq:Eq-IROAM-Solution}
				\mu^{R}_{k} & = & \frac{1}{7} \sum^{7}_{d=1} x^{\mathrm{bf},R}_{k} (S^{\prime}_d)   \\
				\label{eq:Eq-IROAM-Error}
				\sigma^{R}_{k} & = & \sqrt{ \frac{1}{7} \sum^{7}_{d=1} \left( x^{\mathrm{bf},R}_{k} (S^{\prime}_d) - \mu^{R}_{k} \right)^2 }  \,,
			\EA
	\item Compute the parameter-space boundaries where refinement is needed according to
			\BA
				\label{eq:Eq-New-Min-Bnds}
				x^{\mathrm{min},R+1}_k & = & \frac{ \left( \lambda - 1 \right) x^{\mathrm{min},R}_k + \min \left( x^{\mathrm{bf}}_k (S^{\prime}_1) , ..., x^{\mathrm{bf}}_k (S^{\prime}_7) \right) }{\lambda}   \\
				\label{eq:Eq-New-Max-Bnds}
				x^{\mathrm{max},R+1}_k & = & \frac{ \left( \lambda - 1 \right) x^{\mathrm{max},R}_k + \max \left( x^{\mathrm{bf}}_k (S^{\prime}_1) , ..., x^{\mathrm{bf}}_k (S^{\prime}_7) \right) }{\lambda}  \,,
			\EA
			where $\lambda > 1$ is the shrinkage coefficient. It is a user-specified scalar that controls 
			the speed at which the refining region shrinks; larger $\lambda$ values lead {to} slower shrinkage.
	\item Repeat steps 3 through 7 for the refined region until a convergence criterion is met 
			(\eg $\sigma^{R}_{k}$ and/or the refining region are sufficiently small).
\end{enumerate}


\section{Test-case} \label{sec:S-Test-Case}

\subsection{Ground-truth magnetic fields} \label{sec:S-GT-Bfield}

   \begin{deluxetable}{c c c}
   \tablewidth{0.45\textwidth}
   \tablecaption{Ground-truth parameter values
   	\label{tab:Tab-GT-parameters}
	}
   \tablehead{
\colhead{Flux rope model}  & \colhead{$\Phi_{\mathrm{GT}}$ (Mx)} & \colhead{$\mathrm{F}_{\mathrm{GT}}$ (Mx cm$^{-1}$)}
   }
   \tablecomments{Each model is used to produce a set of coronal polarimetric observations to be fitted, as if we were applying the DOCFM to 9 different cavity systems in observational applications. Such a choice allows us to show that the robustness and success of the DOCFM framework are independent of the choice of parameter values for the ground truth 3D magnetic field.}
   \startdata
	  Low-height Low-twist (LL) & $2.50 \times 10^{20}$ & $-1.00 \times 10^{8}$ \\
	  Low-height Mild-twist (LM) & $2.50 \times 10^{20}$ & $-4.00 \times 10^{9}$ \\
	  Low-height High-twist (LH) & $2.50 \times 10^{20}$ & $-1.00 \times 10^{10}$ \\
   	  & & \\
	  Mid-height Low-twist (ML) & $5.00 \times 10^{20}$ & $-1.00 \times 10^{8}$ \\
	  Mid-height Mild-twist (MM) & $5.00 \times 10^{20}$ & $-4.00 \times 10^{9}$ \\
	  Mid-height High-twist (MH) & $5.00 \times 10^{20}$ & $-1.00 \times 10^{10}$ \\
   	  & & \\
	  High-height Low-twist (HL) & $7.50 \times 10^{20}$ & $-1.00 \times 10^{8}$ \\
	  High-height Mild-twist (HM) & $7.50 \times 10^{20}$ & $-4.00 \times 10^{9}$ \\
	  High-height High-twist (HH) & $7.50 \times 10^{20}$ & $-1.00 \times 10^{10}$
   \enddata
   \end{deluxetable} 

  \begin{figure}
   \centerline{\includegraphics[width=0.48\textwidth,clip=]{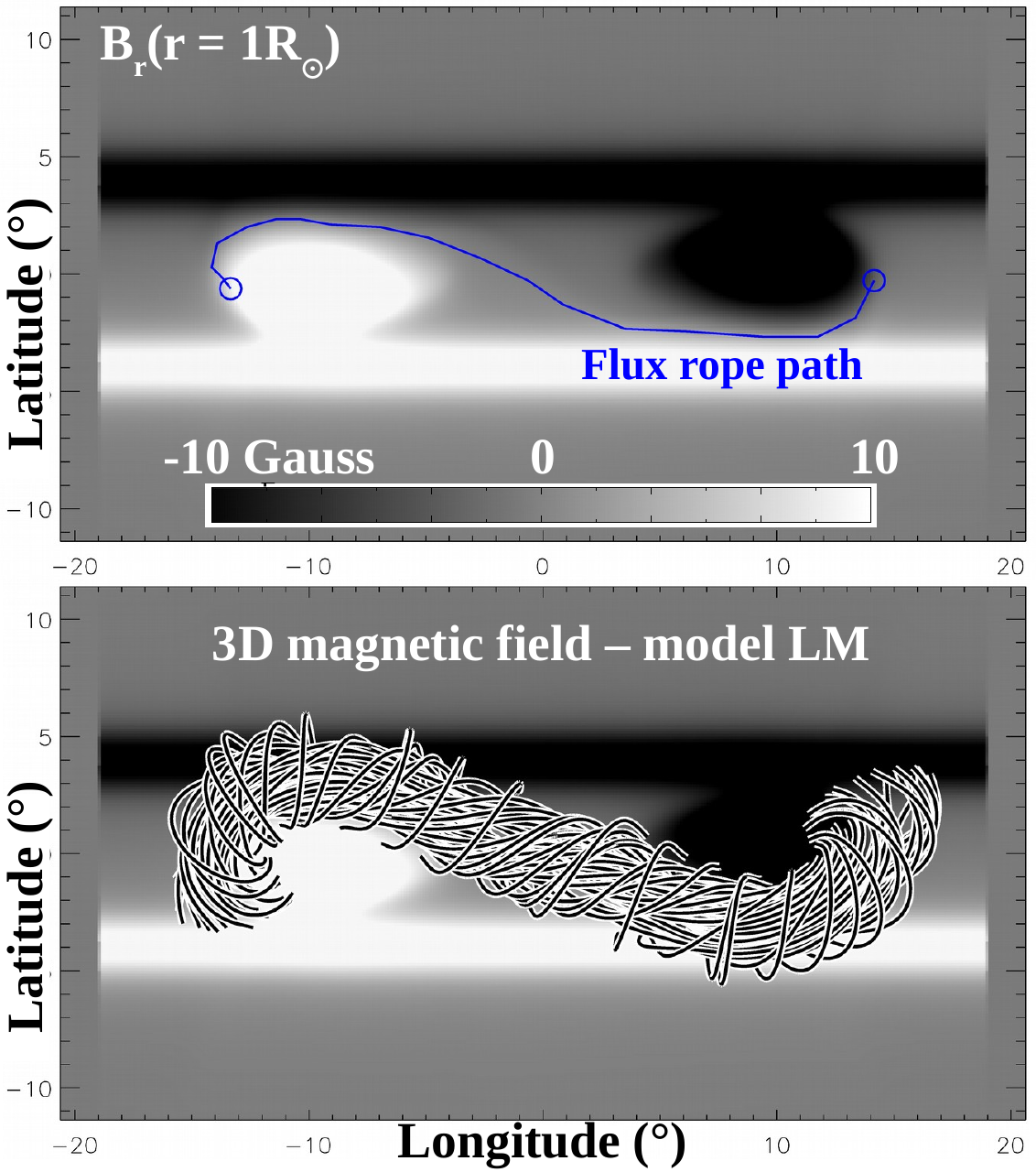}
              }
   \caption{{\bf Top:} Photospheric magnetic flux distribution (grey scale) and flux rope path (blue line) and feet (blue circles) used to prove the concept of the DOCFM approach with the flux rope insertion method, as given from \cite{Fan12} MHD simulation. The magnetic field is in units of Gauss. {\bf Bottom:} Top view of selected 3D magnetic field lines for ground-truth model LM which parameter values are reported in \tab{Tab-GT-parameters}.}
              \label{fig:Fig-GT-Bfield}
   \end{figure} 

%
%
With this series of papers, our goal is to investigate the possibility of improving 
the reliability of 3D magnetic field reconstructions by further exploiting coronal 
polarimetry. The primary goal of this investigation is to prove the concept of 
the methodology and characterize the uncertainties that are both inherent to the limited 
amount of information contained in the polarimetric data (see \sect{S-Introduction}) 
and the limitations of the optimization algorithm (described in \sect{S-IROAM}). 
To that end, we begin by creating a ``ground truth'' solution, which is a full 
representation of the coronal physical state (magnetic field, density, temperature, 
pressure) created from a generative model. With this ground truth solution, we can 
forward model synthetic coronal data that will be used in the place of true observations. 
The point of such an approach is to test the capability of the DOCFM framework for a problem 
where the solution is fully known.

Limitations due to the ability of the generative model to reproduce a given 
magnetic field will be addressed in our second paper in this series (Paper II). 
For these reasons, the ground-truth 
magnetic fields we use in this paper are created with the reconstruction model 
we use, \ie the flux rope insertion method briefly described in \sect{S-Generative-Model}.

%
%
To produce a ground-truth magnetic field with the flux rope insertion method, 
we first need a photospheric magnetic flux distribution and a flux rope path 
(see \sect{S-Generative-Model}). \fig{Fig-GT-Bfield} displays the photospheric magnetogram 
and flux rope path selected for our investigation. The boundary condition is chosen to be 
consistent with that of flux-rope MHD model of \cite{Fan12}, which we will use in Paper II 
to further test the robustness of our approach. For consistency, we computed the PFSS 
from the flux distribution displayed in \fig{Fig-GT-Bfield} with the source surface 
set at $\approx 6 R_{\odot}$, which is far enough from the photosphere to allow most 
of the arcade magnetic field to be closed. To create the ground-truth, current-carrying 
magnetic field, we then insert a flux rope with axial flux, $\Phi_{\mathrm{GT}}$, 
and poloidal flux per unit length, $\mathrm{F}_{\mathrm{GT}}$. \fig{Fig-GT-Bfield} 
displays the ground-truth solution defined by the set of parameters, 
$\Phi_{\mathrm{GT}} = 2.5 \times 10^{20} \ \mathrm{Mx}$ and 
$\mathrm{F}_{\mathrm{GT}} = -4 \times 10^{9} \ \mathrm{Mx} \cdot \mathrm{cm}^{-1}$, 
which produces a low-lying, mildly-twisted, left-handed flux rope referred to as 
model LM.

%
%
To show that the DOCFM approach works regardless of the choice of ground-truth 
parameters, $\left( \Phi_{\mathrm{GT}}, \mathrm{F}_{\mathrm{GT}} \right)$, and validate 
it, we consider 8 additional ground truth solutions that are reported in \tab{Tab-GT-parameters}. 
Each one of the 9 chosen flux ropes is used to create a set of coronal polarimetric observations 
to be fitted, as if we were applying the DOCFM framework to finding the 3D magnetic field 
of 9 different solar coronal cavities. In particular, we produce low-height, mid-height 
and high-height ground truth flux ropes with different degrees of magnetic twist by choosing 
different axial and poloidal flux values.

  \begin{figure*}
   \centerline{\includegraphics[width=0.98\textwidth,clip=]{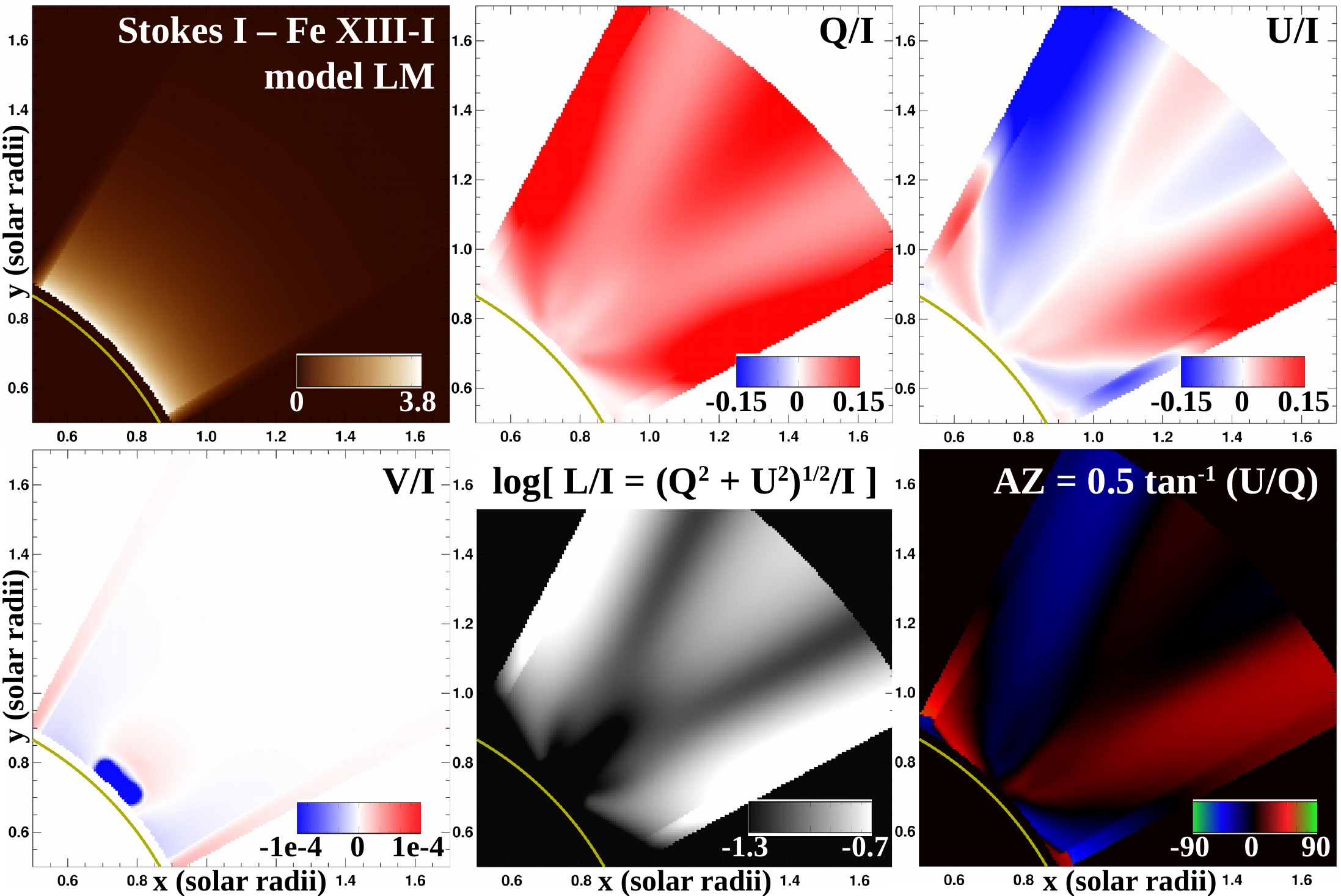}
              }
   \caption{Coronal polarimetric observations in the Fe XIII-I line synthesized with FORWARD for the ground-truth model LM which parameter values are reported in \tab{Tab-GT-parameters}. The flux rope axis is aligned with the LOS. The total line intensity (Stokes $I$), the two components of the fraction of linear polarization ($Q/I$ and $U/I$), and the fraction of circular polarization ($V/I$) are displayed together with the fraction of linear ($L/I = \sqrt{Q^2 + U^2}/I$) polarization and the azimuth ($AZ = 0.5 \tan^{-1} (U/Q)$). Notice the darkening in the central part of the flux rope and the dark lagomorphic shape in the $L/I$ signal, which are typical signatures expected in the $L/I$ signal associated with flux ropes closely aligned with the LOS \citep[\eg][]{BakSteslicka13,Rachmeler13}.}
              \label{fig:Fig-GT-data}
   \end{figure*} 


\subsection{Synthetic data} \label{sec:S-Synthetic-Data}

%
%
Our application of the DOCFM framework has a focus on the exploitation of off-limb 
coronal polarimetry as obtained by the CoMP instrument in the Fe XIII-I line 
\citep[10747 \AA;][]{Tomczyk08}. For our analysis, we use Stokes {\it I}, {\it Q}, 
{\it U} and {\it V} images synthesized with FORWARD (see \sect{S-FORWARD}). 
We further considered Stokes $L = \sqrt{Q^2 + U^2}$ and Stokes $AZ = 0.5 \tan^{-1} (U/Q)$ 
images, which provide an alternate representation of the linear polarization and have been 
shown to provide useful diagnostics of the coronal magnetic field 
\citep[\eg][]{BakSteslicka13,Rachmeler14,Gibson17,KarnaSub}. Each synthetic Stokes image 
is generated with a field of view (FOV) set to $y \times z = [0.5 R_{\odot},1.7 R_{\odot}]^2$ 
($y$ and $z$ being the plane-of-sky coordinates) and $x = [-0.79, 0.79] R_{\odot}$ 
in the LOS direction. The POS and the LOS are respectively covered with $192^2$ and 161 
grid points. The resulting spatial resolution is $\approx 6.0'' \times 6.0''$ in the POS, 
and $\approx 9.4''$ along the LOS. The forward calculations of the Stokes parameters are 
limited to a radial range of $[1.03,1.7] R_{\odot}$, where 1.03 corresponds to the lower 
limit of CoMP FOV. Our choice of spatial resolution of $6.0''$ is only slightly 
{lower} 
than that of CoMP ($4.5''$) to allow us to maintain a relatively low computational time 
per set of Stokes parameters (about 15 minutes on a Linux workstation with an Intel Xeon 
E5-2630 v4 processor).

%
%
Forward modeling of coronal polarimetric observables requires 
a 3D magnetic field model and also a plasma model, because the density and 
temperature are involved in the formation process of any emission line. However, 
most NLFFF generative models, such as the flux rope insertion one, disregard 
the plasma and only provide a 3D magnetic field solution. In the DOCFM framework, 
we therefore need to provide a plasma model if the magnetic field reconstruction 
model does not include one. For a FFF, the plasma solution compatible with 
the force-free assumption is a spherically symmetric hydrostatic atmosphere. 
If we further assume that the atmosphere is isothermal with coronal temperature, 
$T_{c}$, then the plasma density and pressure are
\BA
	\label{eq:Eq-density}
	n(r) & = & n_{c} \exp \left( \frac{R_{c}}{h} \cdot \frac{R_{c} - r}{r} \right)     \\
	\label{eq:Eq-scale-height}
	h & = & \frac{2 k_B T_{c} R^{2}_{c}}{G M_{c} m_p}   \\		
	\label{eq:Eq-pressure}
	P(r) & = & 2 n(r) k_B T_c     \,,
\EA
where $r$ is the radial distance to the center of the Sun (in cm), $R_{c}$ is 
the radius of the solar coronal base, $n_{c} = n(R_{c})$ is the plasma density 
at the coronal base (in units of cm$^{-3}$), $n(r)$ is the coronal plasma density 
(in units of cm$^{-3}$), $h$ is the scale height (in cm), $k_{B}=1.38 \times 10^{-16}$ erg K$^{-1}$ 
is Boltzmann constant (in CGS units), $G = 6.67 \times 10^{-8}$ cm$^3$ g$^{-1}$ s$^{-2}$ 
is the gravitational constant (CGS units), $M_{c}$ is the solar plasma mass at $R_{c}$ 
(in g), $m_p = 1.67 \times 10^{-24}$ g is the proton mass, and $P(r)$ is the plasma 
pressure (in dyne cm$^{-2}$).

\eqss{Eq-density}{Eq-pressure} define the plasma model we specify with each magnetic 
field model we compute in this paper. Assuming that the coronal base is at $\sim$2 Mm 
above the photosphere, then we have $R_{c} \approx R_{\odot} = 6.96 \times 10^{10}$ cm 
(the solar radius) and $M_{c} \approx M_{\odot} = 1.99 \times 10^{33}$ g (the solar 
mass). We set $T_{c} = 1.46 \times 10^{6}$ K and $n_{c} = 1.82 \times 10^8$ cm$^{-3}$ 
in accordance with the values observed at the base of the 3D MHD simulation of 
\cite{Fan12} for consistency with Paper II (cf. \sect{S-GT-Bfield}). 
The resulting coronal polarimetric 
observations for the ground-truth model LM described \sect{S-GT-Bfield} are 
shown in \fig{Fig-GT-data}.


\subsection{Mean squared error diagnostics} \label{sec:S-MSEs}

%
%
In the DOCFM framework, the 3D magnetic field solution is obtained by minimization 
of an MSE between predicted and real polarization signals. Let Y be a Stokes-related 
image, \ie Y can be any of $\{ Q/I, U/I, V/I, L/I, AZ \}$. Working with images of 
polarization fraction is here motivated by the fact that Stokes {\it I}, {\it Q}, 
{\it U} and {\it V} all have dependencies on the plasma density \citep[\eg][]{Casini99} 
and that working with ratios of these quantities reduces the sensitivity to the plasma 
density. $\mathrm{Y}^{\mathrm{GT}}$ is a Stokes-related image associated with the ground-truth 
magnetic field, \ie any of the images shown in \fig{Fig-GT-data}. For the $i$-th vector 
parameter, $\xx^i$, the mean squared error, $\chi^{2}_{\mathrm{Y}} ( \xx^i )$, between 
the predicted (Y$( \xx^i )$) and ground-truth images, is
\BE
	\label{eq:Eq-MSE-image}
	\chi^{2}_{\mathrm{Y}} ( \xx^i ) = \sum_{l} \left( \mathrm{Y}_l ( \xx^i ) - \mathrm{Y}^{\mathrm{GT}}_l \right)^2 \,,
\EE
where $l$ is the $l$-th pixel of the Y image. The final MSEs we consider are then
\BA
	\label{eq:Eq-MSE-QUV}
	\chi^{2}_{QUV} ( \xx^i ) & = & w^{ }_{Q/I} \chi^{2}_{Q/I} ( \xx^i ) + w^{ }_{U/I} \chi^{2}_{U/I} ( \xx^i ) + w^{ }_{V/I} \chi^{2}_{V/I} ( \xx^i )     \\
	\label{eq:Eq-MSE-QU}
	\chi^{2}_{QU} ( \xx^i ) & = & w^{ }_{Q/I} \chi^{2}_{Q/I} ( \xx^i ) + w^{ }_{U/I} \chi^{2}_{U/I} ( \xx^i )    \\		
	\label{eq:Eq-MSE-LAZV}
	\chi^{2}_{LAZV} ( \xx^i ) & = & w^{ }_{L/I} \chi^{2}_{L/I} ( \xx^i ) + w^{ }_{AZ} \chi^{2}_{AZ} ( \xx^i ) + w^{ }_{V/I} \chi^{2}_{V/I} ( \xx^i )    \\
	\label{eq:Eq-MSE-LAZ}
	\chi^{2}_{LAZ} ( \xx^i ) & = & w^{ }_{L/I} \chi^{2}_{L/I} ( \xx^i ) + w^{ }_{AZ} \chi^{2}_{AZ} ( \xx^i )    \,,
\EA
where the $w_{\mathrm{Y}}$ coefficients are used to force the individual MSEs to similarly 
contribute to the overall MSE. Such a choice is motivated by the fact that, $Q/I$, $U/I$, $V/I$, 
$L/I$, and $AZ$ vary on very different scales, \ie $\sim 10^{-2}$ for $(Q/I, U/I, L/I)$, 
$\sim 10^{-4}$ for $V/I$, and $\sim 1 - 10^2$ for $AZ$ \citep[see \eg][]{Judge06,Rachmeler13,Gibson17}. 
By tuning the $w_{\mathrm{Y}}$ coefficients, we can make sure that the overall MSE is 
sensitive to the individual MSE possessing the smallest scale and not dominated by the one 
with the largest scale. The values used for our analysis are 
$( w_{Q/I}, w_{U/I}, w_{L/I}, w_{AZ}, w_{V/I} ) = ( 5.4 \times 10^{-2}, 6.1 \times 10^{-2}, 4.5 \times 10^{-2}, 1.7 \times 10^{-3}, 7.5 \times 10^{3} )$.

%
%
CoMP is currently the only instrument realizing daily observations of coronal emission line 
polarization in the IR Fe XIII lines. Unfortunately, the signal-to-noise ratio for the circular 
polarization (Stokes $V$) is too small for CoMP to allow its routine measurement. Although 
this limitation should be resolved by DKIST or the proposed COSMO telescope \citep{Tomczyk16}, 
we use the $\chi^{2}_{QU}$ MSE to test whether 
the linear polarization signal associated with the Fe XIII lines contains sufficient information 
to fully constrain the coronal magnetic field in our model-data fitting approach. Finally, 
although observations of the linear polarization actually provide Stokes $Q$ and $U$, 
insights at the actual 3D coronal magnetic field are better obtained from the visual inspection 
of Stokes $L$ and $AZ$. The MSEs with Stokes $L$ and $AZ$ are chosen to investigate which 
of $(L, AZ)$ or $(Q, U)$ provides the most useful numerical constraints in our application 
of the DOCFM framework.


\subsection{Error analysis} \label{sec:S-Errors}

%
%
To characterize the uncertainties in our model-data fitting approach of the 3D reconstruction 
of solar coronal magnetic fields, we compute the following errors
\BA
	\label{eq:Eq-Error-Axial}
	\epsilon_{\Phi} & = & \bigg| \frac{ \Phi_{\mathrm{BF}} - \Phi_{\mathrm{GT}} }{ \Phi_{\mathrm{GT}} } \bigg| \\
	 & & \nonumber \\
	 & & \nonumber \\
	\label{eq:Eq-Error-Poloidal}
	\epsilon_{\mathrm{F}} & = & \bigg| \frac{ \mathrm{F}_{\mathrm{BF}} - \mathrm{F}_{\mathrm{GT}} }{ \mathrm{F}_{\mathrm{GT}} } \bigg| \\
	 & & \nonumber \\
	 & & \nonumber \\
	\label{eq:Eq-Error-Bstrength}
	\epsilon_{B} & = & \sqrt{ \frac{1}{N} \sum_{m} \left( \frac{ \| \bb^{\mathrm{BF}}_m \| - \| \bb^{\mathrm{GT}}_m \| }{ \| \bb^{\mathrm{GT}}_m \| } \right)^2 }  \\
	 & & \nonumber \\
	 & & \nonumber \\
	\label{eq:Eq-Error-Bangle}
	\epsilon_{B\mathrm{-angle}} & = & \sqrt{ \frac{1}{N} \sum_{m} \theta^{2}_{m} }  \\
	 & & \nonumber \\
	 & & \nonumber \\
	\label{eq:Eq-Error-BCWL}
	\epsilon_{B\mathrm{-CWL}} & = & \sin^{-1} \left( \frac{ \sum_{m} \| \bb^{\mathrm{BF}}_m \| \sin \left( \theta_m \right) }{ \sum_{m} \| \bb^{\mathrm{BF}}_m \| } \right) \\
	 & & \nonumber \\
	 & & \nonumber \\
	\label{eq:Eq-Error-Jstrength}
	\epsilon_{J} & = & \sqrt{ \frac{1}{N} \sum_{i} \left( \frac{ \| \jj^{\mathrm{BF}}_m \| - \| \jj^{\mathrm{GT}}_m \| }{ \| \jj^{\mathrm{GT}}_m \| } \right)^2 }  \\
	 & & \nonumber \\
	 & & \nonumber \\
	\label{eq:Eq-Error-Jangle}
	\epsilon_{J\mathrm{-angle}} & = & \sqrt{ \frac{1}{N} \sum_{m} \alpha^{2}_{m} }  \\
	 & & \nonumber \\
	 & & \nonumber \\
	\label{eq:Eq-Error-JCWL}
	\epsilon_{J\mathrm{-CWL}} & = & \sin^{-1} \left( \frac{ \sum_{m} \| \jj^{\mathrm{BF}}_m \| \sin \left( \alpha_m \right) }{ \sum_{m} \| \jj^{\mathrm{BF}}_m \| } \right) \\
	 & & \nonumber \\
	 & & \nonumber \\
	\label{eq:Eq-Error-Energy}
	\epsilon_{E_{\mathrm{free}}} & = & \bigg| \frac{ E_{\mathrm{free}} (\bb_{\mathrm{BF}}) - E_{\mathrm{free}} (\bb_{\mathrm{GT}}) }{ E_{\mathrm{free}} (\bb_{\mathrm{GT}}) } \bigg| \\
	 & & \nonumber \\
	 & & \nonumber \\
	\label{eq:Eq-Error-Helicity}
	\epsilon_{H_r} & = & \bigg| \frac{ H_r (\bb_{\mathrm{BF}}) - H_r (\bb_{\mathrm{GT}}) }{ H_r (\bb_{\mathrm{GT}}) } \bigg|  \\
	 & & \nonumber \\
	 & & \nonumber \\
	\label{eq:Eq-Error-StokesY}
	\epsilon_{\mathrm{Y}} & = & \sqrt{ \frac{1}{N'} \sum_{l} \left( \mathrm{Y}^{\mathrm{BF}}_l - \mathrm{Y}^{\mathrm{GT}}_l \right)^2 }  \,,
\EA
where the subscript $m$ runs over the grid points of the 3D magnetic field computational 
domain, $N$ is the number of grid points, $E_{\mathrm{free}}$ and $H_r$ respectively are 
the free magnetic energy and the relative magnetic helicity as computed in \cite{Bobra08}, 
{the} 
subscript $l$ runs over the pixels of the synthetic Stokes image $\mathrm{Y} = \{ I, Q, U, V \}$, 
$N'$ is the number of pixels in the synthetic Stokes image, and
\BA
	\theta_m & = & \sin^{-1} \left( \frac{ \| \bb^{\mathrm{BF}}_m \times \bb^{\mathrm{GT}}_m \| }{ \| \bb^{\mathrm{BF}}_m \| \ \| \bb^{\mathrm{GT}}_m \| } \right) \\
	 & & \nonumber \\
	 & & \nonumber \\
	\alpha_m & = & \sin^{-1} \left( \frac{ \| \jj^{\mathrm{BF}}_m \times \jj^{\mathrm{GT}}_m \| }{ \| \jj^{\mathrm{BF}}_m \| \ \| \jj^{\mathrm{GT}}_m \| } \right) \,.
\EA

$\epsilon_{\Phi}$, $\epsilon_{\mathrm{F}}$, $\epsilon_{B}$, $\epsilon_{J}$, 
$\epsilon_{E_{\mathrm{free}}}$ and $\epsilon_{H_r}$ are relative errors. $\epsilon_{\mathrm{Y}}$ 
is in units of parts per million (ppm). $\epsilon_{B\mathrm{-angle}}$, 
$\epsilon_{B\mathrm{-CWL}}$, $\epsilon_{J\mathrm{-angle}}$, and $\epsilon_{J\mathrm{-CWL}}$ 
are angles in units of degrees. Note that $\epsilon_{B\mathrm{-CWL}}$ and $\epsilon_{J\mathrm{-CWL}}$ 
are CW-like (CWL) angles defined in the spirit of Equations (13) and (14) of \cite{Wheatland00}.


\section{Results} \label{sec:S-Results}

\subsection{Optimization results} \label{sec:S-Optimization-Results}

%
%
To apply the DOCFM approach, we fix the un-refined starting parameter space grid to 
{$\Phi (\mathrm{Mx}) \times \mathrm{F} (\mathrm{Mx \ cm}^{-1}) = [10^{20}, 10^{21}] \times [-5 \times 10^{10}, +5 \times 10^{10}]$.} 
Four models with axial fluxes spanning four orders of magnitude ($10^{19}$, $10^{20}$, 
$10^{21}$, $10^{22}$) were first run to restrict the axial flux, $\Phi$, to a range of values 
always strong enough to produce a flux rope above the polarity inversion line, and 
low enough to ensure that the flux rope insertion (\sect{S-Generative-Model}) can lead 
to a stable flux rope (stable in the sense that the flux rope does not erupt during 
the magnetofrictional relaxation stage; cf. \sect{S-Generative-Model}). 
For the optimization steps 
with ROAM, we use 3 LHS of 24 points at each iteration, resulting in the computation of 72 flux 
rope models per iteration. For each iteration, the 72 flux rope models are generated (including 
the magnetofrictional relaxation phase; see \sect{S-Generative-Model}) in parallel using the 
{high-performance computing (HPC)} 
resources of the CALMIP\footnote{https://www.calmip.univ-toulouse.fr/} supercomputing center, 
which provides us with the advantage that one iteration has an elapsed time equivalent 
to the generation of one flux rope model only. For the computation of the new boundaries 
where refinement is to be done at each iteration (see \eqs{Eq-New-Min-Bnds}{Eq-New-Max-Bnds}), 
we set the shrinkage coefficient to $\lambda = 2$.

   \begin{deluxetable}{c c c c c}
   \tablewidth{0.48\textwidth}
   \tablecaption{Optimization results vs. iteration number\\ for ground-truth model LM with $\chi^{2}_{QUV}$ minimization
   	\label{tab:Tab-Optimization-vs-Iteration}
	}
   \tablehead{
$R$ & \colhead{$\Phi_{\mathrm{BF}}$ (Mx)} & \colhead{$\sigma_{\Phi}$ (Mx)} & \colhead{$\mathrm{F}_{\mathrm{BF}}$ (Mx cm$^{-1}$)} & \colhead{$\sigma_{\mathrm{F}}$ (Mx cm$^{-1}$)}
   }
   \tablecomments{$\mu = (\Phi_{\mathrm{BF}}, \mathrm{F}_{\mathrm{BF}})$ and $\sigma = (\sigma_{\Phi}, \sigma_{\mathrm{F}})$ are the mean and standard deviation values computed from the seven best-fit vector parameters obtained at each iteration of ROAM, as defined by \eqs{Eq-IROAM-Solution}{Eq-IROAM-Error} in \sect{S-IROAM}. The ground-truth model LM is defined by $(\Phi_{\mathrm{GT}}, \mathrm{F}_{\mathrm{GT}}) = (2.50 \times 10^{20} \ \mathrm{Mx}, \ -4.00 \times 10^{9} \ \mathrm{Mx} \ \mathrm{cm}^{-1})$.}
   \startdata
	  0 & $2.55 \times 10^{20}$ & $2.58 \times 10^{19}$ & $-5.53 \times 10^{8}$ & $1.73 \times 10^{10}$ \\
	  1 & $2.51 \times 10^{20}$ & $3.66 \times 10^{18}$ & $-3.93 \times 10^{9}$ & $2.84 \times 10^{9}$ \\
	  2 & $2.51 \times 10^{20}$ & $8.09 \times 10^{17}$ & $-3.65 \times 10^{9}$ & $5.61 \times 10^{8}$ \\
	  3 & $2.50 \times 10^{20}$ & $2.46 \times 10^{17}$ & $-4.00 \times 10^{9}$ & $2.18 \times 10^{8}$
   \enddata
   \end{deluxetable} 

  \begin{figure}
   \centerline{\includegraphics[width=0.49\textwidth,clip=]{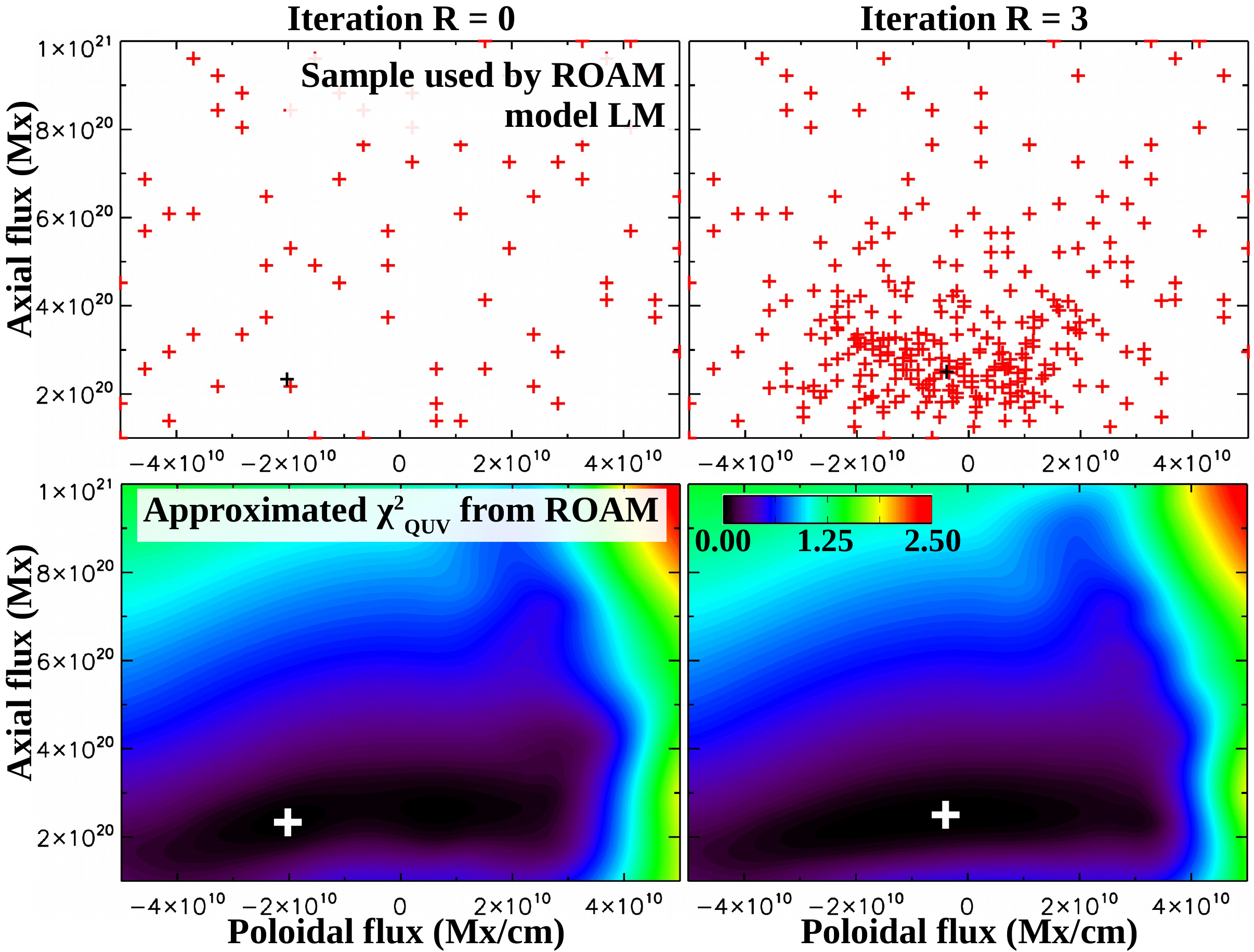}
              }
   \caption{Optimization results for sample $S^{\prime}_{7}$ (see \sect{S-IROAM}) at the un-refined starting stage (left column) and at the refined final stage (right column) for model LM (see \tab{Tab-GT-parameters} for its parameter values). {\bf Top:} Distribution of models in the parameter space (red ``$+$'' signs). {\bf Bottom:} Corresponding approximated $\chi^{2}_{QUV}$ MSE (defined by \eq{Eq-MSE-QUV}). The black {(top row)} / white {(bottom row)} ``$+$'' sign indicates the {position of the minimum} found by ROAM.}
              \label{fig:Fig-Optimization-vs-Iterations}
   \end{figure} 

   \begin{deluxetable}{c c c c c}
   \tablewidth{0.48\textwidth}
   \tablecaption{Optimization results for all mean squared error tests\\ for ground-truth model LM
   	\label{tab:Tab-Optimization-vs-MSE}
	}
   \tablehead{
Minimized & \colhead{$\Phi_{\mathrm{BF}}$} & \colhead{$\sigma_{\Phi}$} & \colhead{$\mathrm{F}_{\mathrm{BF}}$} & \colhead{$\sigma_{\mathrm{F}}$}\\
MSE & (Mx) & (Mx) & (Mx cm$^{-1}$) & (Mx cm$^{-1}$)
   }
   \tablecomments{$\mu = (\Phi_{\mathrm{BF}}, \mathrm{F}_{\mathrm{BF}})$ and $\sigma = (\sigma_{\Phi}, \sigma_{\mathrm{F}})$ are the mean and standard deviation values computed from the seven best-fit vector parameters obtained after $R=3$ iterations of ROAM, as defined by \eqs{Eq-IROAM-Solution}{Eq-IROAM-Error} in \sect{S-IROAM}. The ground-truth model LM is defined by $(\Phi_{\mathrm{GT}}, \mathrm{F}_{\mathrm{GT}}) = (2.50 \times 10^{20} \ \mathrm{Mx}, \ -4.00 \times 10^{9} \ \mathrm{Mx} \ \mathrm{cm}^{-1})$.}
   \startdata
	  $\chi^{2}_{QUV}$ & $2.50 \times 10^{20}$ & $2.46 \times 10^{17}$ & $-4.00 \times 10^{9}$ & $2.18 \times 10^{8}$ \\
	  $\chi^{2}_{QU}$ & $2.50 \times 10^{20}$ & $1.32 \times 10^{17}$ & $-3.93 \times 10^{9}$ & $8.85 \times 10^{7}$ \\
	  $\chi^{2}_{LAZV}$ & $2.50 \times 10^{20}$ & $2.17 \times 10^{17}$ & $-4.00 \times 10^{9}$ & $1.85 \times 10^{8}$ \\
	  $\chi^{2}_{LAZ}$ & $2.50 \times 10^{20}$ & $9.12 \times 10^{16}$ & $-3.98 \times 10^{9}$ & $4.75 \times 10^{7}$
   \enddata
   \end{deluxetable} 

   \begin{deluxetable*}{c c c c c c c c}
   \tablewidth{0.95\textwidth}
   \tablecaption{Optimization results vs. ground-truth parameters for $\chi^{2}_{QUV}$ minimization
   	\label{tab:Tab-Optimization-results-vs-GTs}
	}
   \tablehead{
\colhead{Flux rope model}  & \colhead{$\Phi_{\mathrm{GT}}$ (Mx)}& \colhead{$\Phi_{\mathrm{BF}}$ (Mx)} & \colhead{$\sigma_{\Phi}$ (Mx)} & \colhead{$\mathrm{F}_{\mathrm{GT}}$ (Mx cm$^{-1}$)} & \colhead{$\mathrm{F}_{\mathrm{BF}}$ (Mx cm$^{-1}$)} & \colhead{$\sigma_{\mathrm{F}}$ (Mx cm$^{-1}$)} & \colhead{$R_f$}
   }
   \tablecomments{$\mu = (\Phi_{\mathrm{BF}}, \mathrm{F}_{\mathrm{BF}})$ and $\sigma = (\sigma_{\Phi}, \sigma_{\mathrm{F}})$ are the mean and standard deviation values computed from the seven best-fit vector parameters obtained at the final iteration of ROAM, $R_f$, as defined by \eqs{Eq-IROAM-Solution}{Eq-IROAM-Error} in \sect{S-IROAM}.}
   \startdata
	  LL & $2.50 \times 10^{20}$ & $2.50 \times 10^{20}$ & $3.46 \times 10^{16}$ & $-1.00 \times 10^{8}$ & $-1.08 \times 10^{8}$ & $2.43 \times 10^{7}$ & 5 \\
	  LM & $2.50 \times 10^{20}$ & $2.50 \times 10^{20}$ & $2.46 \times 10^{17}$ & $-4.00 \times 10^{9}$ & $-4.00 \times 10^{9}$ & $2.18 \times 10^{8}$ & 3 \\
	  LH & $2.50 \times 10^{20}$ & $2.50 \times 10^{20}$ & $2.86 \times 10^{17}$ & $-1.00 \times 10^{10}$ & $-0.98 \times 10^{10}$ & $1.35 \times 10^{8}$ & 3 \\
   	  & & & & & & & \\
	  ML & $5.00 \times 10^{20}$ & $5.00 \times 10^{20}$ & $3.46 \times 10^{16}$ & $-1.00 \times 10^{8}$ & $-1.02 \times 10^{8}$ & $6.79 \times 10^{6}$ & 4 \\
	  MM & $5.00 \times 10^{20}$ & $5.00 \times 10^{20}$ & $3.14 \times 10^{16}$ & $-4.00 \times 10^{9}$ & $-4.03 \times 10^{9}$ & $1.21 \times 10^{7}$ & 3 \\
	  MH & $5.00 \times 10^{20}$ & $5.00 \times 10^{20}$ & $7.48 \times 10^{16}$ & $-1.00 \times 10^{10}$ & $-1.00 \times 10^{10}$ & $1.48 \times 10^{7}$ & 3 \\
   	  & & & & & & & \\
	  HL & $7.50 \times 10^{20}$ & $7.50 \times 10^{20}$ & $3.19 \times 10^{17}$ & $-1.00 \times 10^{8}$ & $-0.93 \times 10^{8}$ & $3.05 \times 10^{7}$ & 3 \\
	  HM & $7.50 \times 10^{20}$ & $7.49 \times 10^{20}$ & $6.36 \times 10^{17}$ & $-4.00 \times 10^{9}$ & $-3.95 \times 10^{9}$ & $1.26 \times 10^{7}$ & 3 \\
	  HH & $7.50 \times 10^{20}$ & $7.50 \times 10^{20}$ & $2.24 \times 10^{17}$ & $-1.00 \times 10^{10}$ & $-1.00 \times 10^{10}$ & $4.02 \times 10^{7}$ & 3
   \enddata
   \end{deluxetable*} 

%
%
\tab{Tab-Optimization-vs-Iteration} reports the optimization results obtained when applying 
the DOCFM approach to reconstructing the 3D magnetic field of the flux rope model LM 
from the minimization of the $\chi^{2}_{QUV}$ MSE defined in \eq{Eq-MSE-QUV}. It shows that 
convergence towards a best-fit solution is achieved in only 4 applications (\ie $R = 3$ 
iterations) of the optimization procedure described in \sect{S-IROAM}, which corresponds 
to 288 model evaluations. The best-fit solution, as computed from \eq{Eq-IROAM-Solution}, 
is obtained for $\Phi_{\mathrm{BF}} = 2.50 \times 10^{20} \ \mathrm{Mx} \pm 2.46 \times 10^{17}$ 
and $\mathrm{F}_{\mathrm{BF}} = -4.00 \times 10^{9} \ \mathrm{Mx \ cm}^{-1} \pm 2.18 \times 10^{8}$, 
while the ground-truth parameters for model LM are 
$(\Phi_{\mathrm{GT}}, \mathrm{F}_{\mathrm{GT}}) = (2.50 \times 10^{20} \ \mathrm{Mx}, \ -4.00 \times 10^{9} \ \mathrm{Mx} \ \mathrm{cm}^{-1})$. 
Our model-data fitting approach is therefore able to accurately retrieve the ground-truth 
parameters of the flux rope model LM.

%
%
\tab{Tab-Optimization-vs-Iteration} further details the mean, $\mu = (\Phi_{\mathrm{BF}}, \mathrm{F}_{\mathrm{BF}})$, 
and standard deviation, $\sigma = (\sigma_{\Phi}, \sigma_{\mathrm{F}})$, values computed from the seven best-fit 
vector parameters obtained at each iteration of ROAM, as defined by \eqs{Eq-IROAM-Solution}{Eq-IROAM-Error}. 
The standard deviation value reflects the disagreement between these seven best-fit vector parameters. The table 
shows that the standard deviation tends to display very high values at the first application of the optimization 
procedure and progressively decreases as the number of iteration increases. For the first few iterations, this is 
because the seven best-fit vector parameters can strongly depend on the sparse sample used to apply ROAM 
\citep[\cf][]{Dalmasse16}. As the optimization procedure is iterated, this disagreement decreases and the seven 
best-fit vector parameters eventually converge towards the same solution, which only marginally depends on the sparse 
sample used with ROAM. Therefore, the proposed iterative implementation of ROAM (see \sect{S-IROAM}) makes it more 
robust against the choice of sparse sample used to apply it.

%
%
The results reported in \tab{Tab-Optimization-vs-Iteration} also show that the poloidal flux parameter, 
$\mathrm{F}_{\mathrm{BF}}$, is systematically associated with a much higher standard deviation than the axial flux 
parameter, $\Phi_{\mathrm{BF}}$, and requires more iterations for the seven best-fit vector parameters to converge 
towards a consistent solution. From the analysis of the approximated MSE (shown in \fig{Fig-Optimization-vs-Iterations}), 
we find that this is related with the fact that the minimum region is very flat and extended, and hence poorly 
defined for the poloidal flux as compared with the axial flux. The latter is explained by the fact that, within 
the setup considered in this work, the set of polarimetric data exploited to reconstruct the 3D coronal magnetic 
field is much more sensitive to the axial flux of the flux rope than to its poloidal flux. These polarization 
data thus provide stronger constraints on the former than on the latter. This, however, only applies to the setup 
considered in this paper and would require additional investigations to determine whether it is true regardless 
of, \eg the flux rope axis orientation with regard to the LOS-direction.

%
%
The optimization results for all MSE tests for this 
ground-truth model LM are presented in \tab{Tab-Optimization-vs-MSE}, 
after $R=3$ iterations. At intermediate iteration steps (not shown here), we do not systematically find a one-to-one 
correspondence between the seven best-fit vector parameters obtained from the $\chi^{2}_{LAZV}$-minimization and 
those obtained from the $\chi^{2}_{QUV}$-minimization. This can be explained by the fact that the two datasets bring 
constraints on the 3D coronal magnetic field in different forms, which results in different MSE surfaces and shapes 
that ROAM necessarily approximates differently. However, the spread of these seven best-fit vector parameters remains 
very similar regardless of the iteration number and dataset used. Once convergence is achieved, the differences between 
the solution, $(\Phi_{\mathrm{BF}}, \mathrm{F}_{\mathrm{BF}})$, from the $\chi^{2}_{LAZV}$-minimization and that 
from the $\chi^{2}_{QUV}$-minimization are only marginal. For the tests and setup considered in this paper, we thus do 
not find any clear evidence that performing the optimization with the $\{ L/I, AZ, V/I \}$ dataset provides a particular 
advantage or disadvantage over using the $\{ Q/I, U/I, V/I \}$ dataset. Further investigations with different LHS 
distributions and magnetic field models are required to determine whether this is a general trend.

%
%
\tab{Tab-Optimization-vs-MSE} further compares the optimization results obtained when only the linear polarization 
signal (\ie $\{ L/I, AZ \}$ or $\{ Q/I, U/I \}$) is used with the DOCFM approach. It shows that, once convergence 
is achieved, the solution obtained from minimizing $\chi^{2}_{QU}$ (resp. $\chi^{2}_{LAZ}$) is only marginally 
different from the solution obtained when minimizing $\chi^{2}_{QUV}$ (resp. $\chi^{2}_{LAZV}$). While it does 
not mean that it will always be the case in modeling and/or observational studies, we find that the fraction of 
linear polarization and the azimuth together contain enough information to constrain the 3D coronal magnetic 
field. This is a combination of the van Vleck effect \citep{VanVleck25} and magnetic flux component along 
the LOS. Both produce extinctions in the linear polarization signal that carry information about both the POS 
and LOS magnetic field components 
\citep[as already described in previous studies; see, \eg][]{BakSteslicka13,Rachmeler14,Gibson17,KarnaSub}, 
and hence about the coronal electric current system. Such 
signatures are enough to constrain the original 3D magnetic field and retrieve its ground-truth 
parameters with the DOCFM method.

%
%
Finally, the results of applying the DOCFM approach to all 9 flux rope ground-truth models reported 
in \tab{Tab-GT-parameters} are presented in \tab{Tab-Optimization-results-vs-GTs} for $\chi^{2}_{QUV}$ 
minimization. It shows that our model-data fitting method is able to accurately retrieve the ground-truth 
parameters of each flux rope model, regardless of their values. The analysis of these 9 tests as a function 
of iteration number and dataset used for the $\chi^{2}$ minimization confirms all our previous findings.


\subsection{3D magnetic field comparison} \label{sec:S-B-comparison}

  \begin{figure*}
   \centerline{\includegraphics[width=0.82\textwidth,clip=]{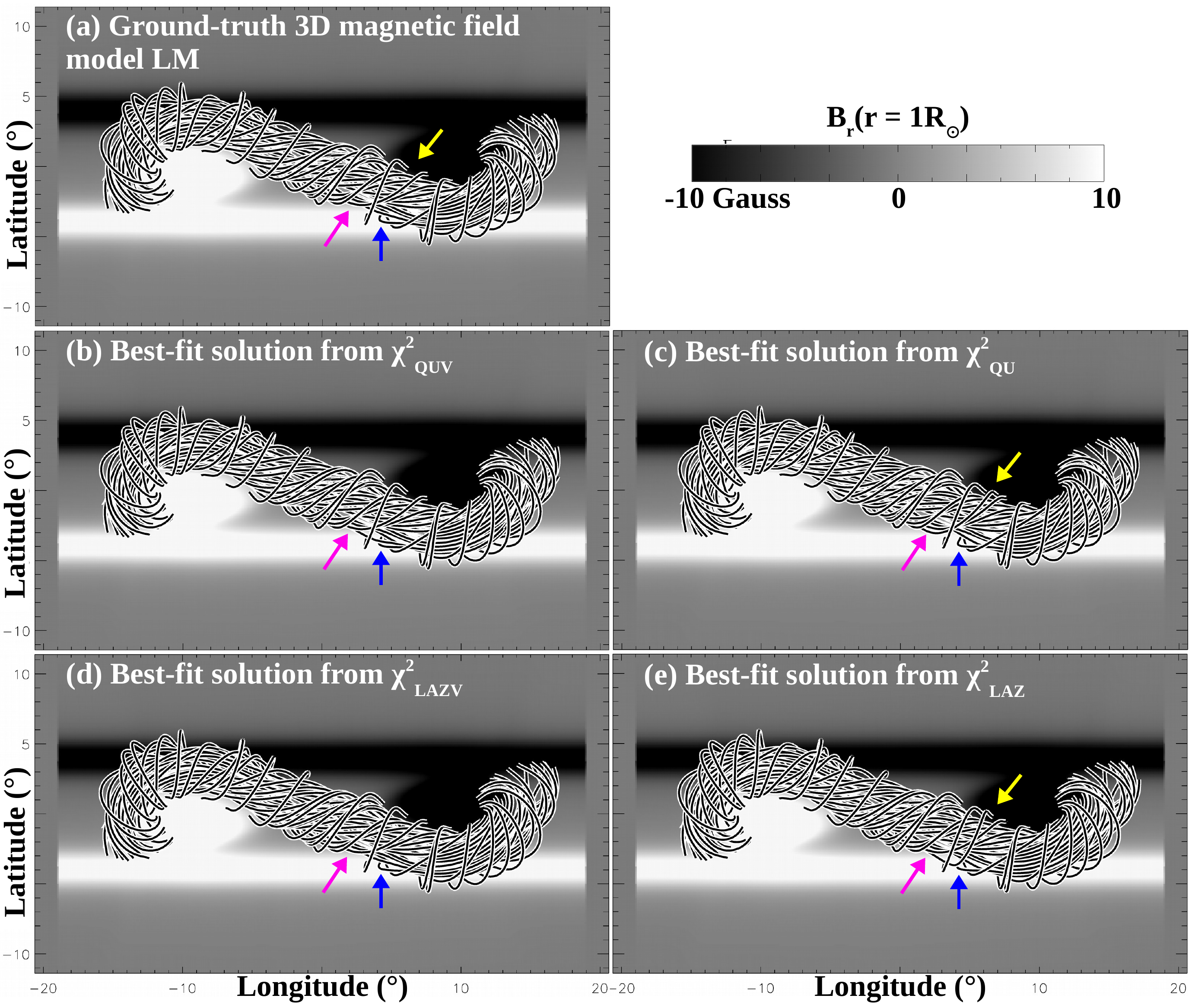}
              }
   \caption{Top view of the selected field-lines from {(a)} the ground-truth magnetic field model LM (see \tab{Tab-GT-parameters} for its parameter values) and {(b -- e)} the optimization solutions (reported in \tab{Tab-Optimization-vs-MSE}) associated with the $\chi^{2}_{QUV}$, $\chi^{2}_{QU}$, $\chi^{2}_{LAZV}$, and $\chi^{2}_{LAZ}$ MSEs defined in \eqss{Eq-MSE-QUV}{Eq-MSE-LAZ}. The field-lines from the DOCFM solutions were all integrated from the same {selected} starting footpoints as the ground-truth magnetic field lines. A few colored arrows have been added to emphasize regions where magnetic field lines are different from the ground-truth magnetic field ones.}
              \label{fig:Fig-Bfields-comparison}
   \end{figure*} 

   \begin{deluxetable}{c c}
   \tablewidth{0.4\textwidth}
   \tablecaption{Best-fits vs. ground-truth 3D magnetic field\\ for flux rope model LM
   	\label{tab:Tab-BF-errors-LM}
	}
   \tablehead{
 \colhead{Error} & \colhead{Upper value from the 4 solutions obtained}  \\
 & \colhead{with $\chi^{2}_{QUV}$, $\chi^{2}_{QU}$, $\chi^{2}_{LAZV}$ and $\chi^{2}_{LAZ}$}
   }
   \tablecomments{All $\epsilon$ are errors defined by \eqss{Eq-Error-Axial}{Eq-Error-StokesY}. $\epsilon_{\Phi}$, $\epsilon_{\mathrm{F}}$, $\epsilon_{B}$, $\epsilon_{J}$, $\epsilon_{E_{\mathrm{free}}}$ and $\epsilon_{H_r}$ are relative errors. $\epsilon_{I}$, $\epsilon_{Q}$, $\epsilon_{U}$ and $\epsilon_{V}$ are Stokes errors in ppm. $\epsilon_{B\mathrm{-angle}}$, $\epsilon_{B\mathrm{-CWL}}$, $\epsilon_{J\mathrm{-angle}}$ and $\epsilon_{J\mathrm{-CWL}}$ are angle errors in degrees. For brevity and because values are very similar regardless of the nature of the minimized MSE, only the largest errors found are displayed.}
   \startdata
	  $\epsilon_{\Phi}$ & $4.47 \times 10^{-4}$  \\
	  $\epsilon_{\mathrm{F}}$ & $1.64 \times 10^{-2}$  \\
	  $\epsilon_{B}$ & $5.34 \times 10^{-4}$ \\
	  $\epsilon_{B\mathrm{-angle}}$ & $2.44 \times 10^{-2}$  \\
	  $\epsilon_{B\mathrm{-CWL}}$ & $4.08 \times 10^{-2}$ \\
	  $\epsilon_{J}$ & $3.13 \times 10^{-2}$   \\
	  $\epsilon_{J\mathrm{-angle}}$ & 1.38  \\
	  $\epsilon_{J\mathrm{-CWL}}$ & $1.93 \times 10^{-1}$ \\
	  $\epsilon_{E_{\mathrm{free}}}$ & $1.38 \times 10^{-3}$ \\
	  $\epsilon_{H_r}$ & $1.09 \times 10^{-3}$ \\
	  $\epsilon_{I}$ & $1.26 \times 10^{-5}$  \\
	  $\epsilon_{Q}$ & $2.40 \times 10^{-5}$  \\
	  $\epsilon_{U}$ & $1.66 \times 10^{-5}$  \\
	  $\epsilon_{V}$ & $1.70 \times 10^{-7}$
   \enddata
   \end{deluxetable} 

   \begin{deluxetable}{c c}
   \tablewidth{0.4\textwidth}
   \tablecaption{Best-fits vs. ground-truth 3D magnetic field\\ for $\chi^{2}_{QUV}$ minimization
   	\label{tab:Tab-BF-errors-all-FRs}
	}
   \tablehead{
 \colhead{Error} & \colhead{Upper value from the solutions obtained}  \\
 & \colhead{for all flux rope types of \tab{Tab-Optimization-results-vs-GTs}}
   }
   \tablecomments{All $\epsilon$ are errors defined by \eqss{Eq-Error-Axial}{Eq-Error-StokesY}. $\epsilon_{\Phi}$, $\epsilon_{\mathrm{F}}$, $\epsilon_{B}$, $\epsilon_{J}$, $\epsilon_{E_{\mathrm{free}}}$ and $\epsilon_{H_r}$ are relative errors. $\epsilon_{I}$, $\epsilon_{Q}$, $\epsilon_{U}$ and $\epsilon_{V}$ are Stokes errors in ppm. $\epsilon_{B\mathrm{-angle}}$, $\epsilon_{B\mathrm{-CWL}}$, $\epsilon_{J\mathrm{-angle}}$ and $\epsilon_{J\mathrm{-CWL}}$ are angle errors in degrees. Results are representative of the errors obtained from minimizing all the other MSEs considered in the paper. For brevity and because values are very similar regardless of the flux rope type, only the largest errors found are displayed.}
   \startdata
	  $\epsilon_{\Phi}$ & $1.41 \times 10^{-3}$  \\
	  $\epsilon_{\mathrm{F}}$ & $7.64 \times 10^{-2}$  \\
	  $\epsilon_{B}$ & $1.19 \times 10^{-3}$ \\
	  $\epsilon_{B\mathrm{-angle}}$ & $5.41 \times 10^{-2}$  \\
	  $\epsilon_{B\mathrm{-CWL}}$ & $9.01 \times 10^{-2}$ \\
	  $\epsilon_{J}$ & $5.52 \times 10^{-2}$   \\
	  $\epsilon_{J\mathrm{-angle}}$ & 1.34  \\
	  $\epsilon_{J\mathrm{-CWL}}$ & $3.41 \times 10^{-1}$ \\
	  $\epsilon_{E_{\mathrm{free}}}$ & $2.88 \times 10^{-3}$ \\
	  $\epsilon_{H_r}$ & $1.99 \times 10^{-3}$ \\
	  $\epsilon_{I}$ & $3.08 \times 10^{-5}$  \\
	  $\epsilon_{Q}$ & $5.98 \times 10^{-5}$  \\
	  $\epsilon_{U}$ & $5.24 \times 10^{-5}$  \\
	  $\epsilon_{V}$ & $4.18 \times 10^{-7}$
   \enddata
   \end{deluxetable} 

  \begin{figure*}
   \centerline{\includegraphics[width=0.98\textwidth,clip=]{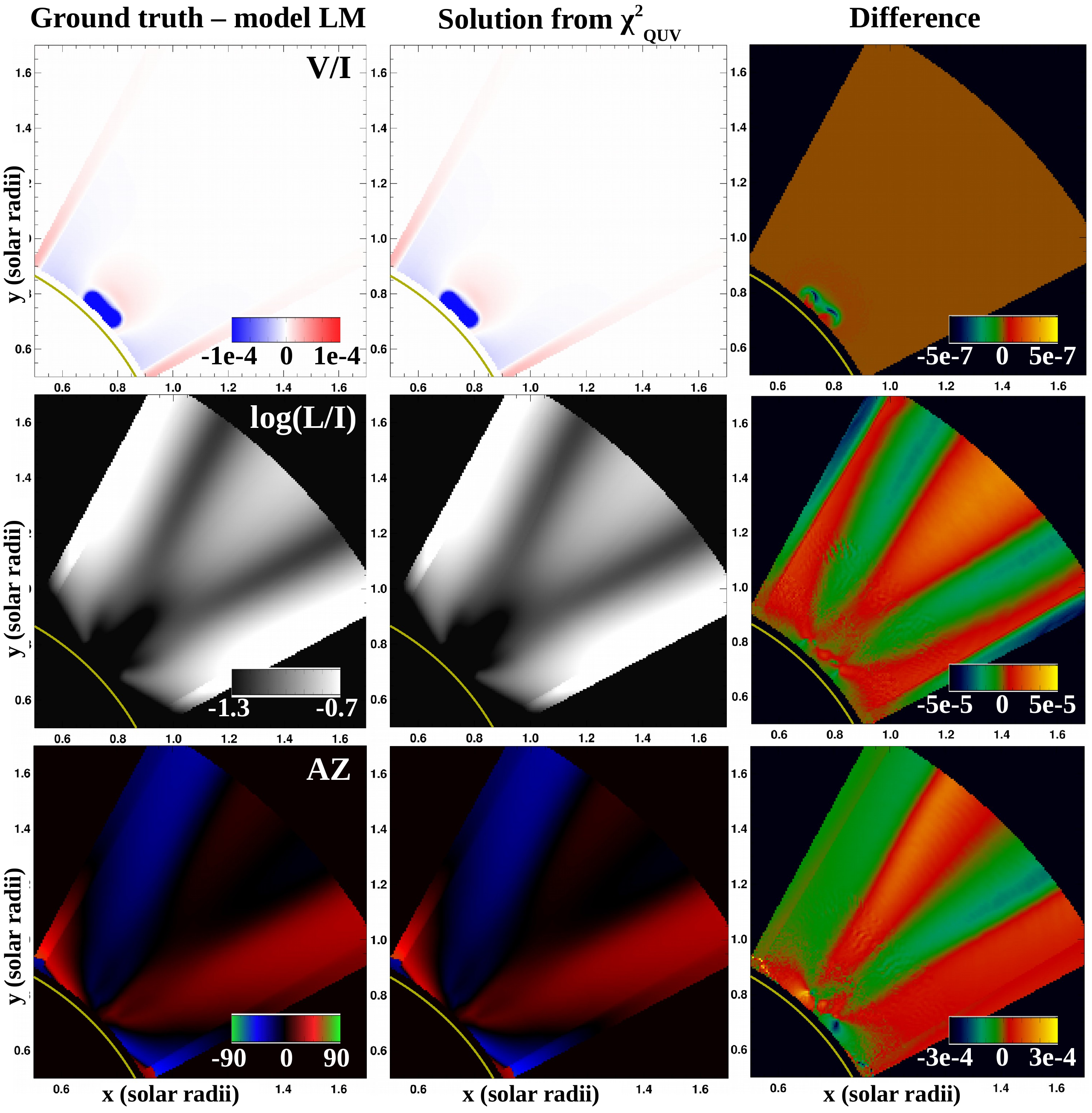}
              }
   \caption{Comparison between Stokes-related images for the ground-truth magnetic field model LM (see \tab{Tab-GT-parameters} for its parameter values) and for the solution obtained from the $\chi^{2}_{QUV}$ MSE defined in \eq{Eq-MSE-QUV}.}
              \label{fig:Fig-Stokes-comparison}
   \end{figure*} 

%
%
We now focus on the qualitative and quantitative comparison of the 3D magnetic field solution obtained 
by the DOCFM method and the actual ground-truth magnetic field. \fig{Fig-Bfields-comparison} displays 
the 3D magnetic field lines for the flux rope model LM 
{(panel (a))} 
and the DOCFM solution 
({panels (b) to (e), which parameter values are} 
reported in \tab{Tab-Optimization-vs-MSE}) 
obtained from each $\chi^2$ minimization defined by \eqss{Eq-MSE-QUV}{Eq-MSE-LAZ}. 
{The flux rope field-lines from the DOCFM solutions were all integrated from the same selected 
starting footpoints as the ground-truth magnetic field lines.} 
As one can see from 
the colored arrows, differences exist between the DOCFM solution and the ground-truth magnetic field lines 
{(\ie compare panels (b -- e) to panel (a))}. 
However, these differences are very small and the 3D magnetic field is -- qualitatively -- very 
well recovered, regardless of the dataset used to perform the $\chi^2$ minimization. The latter is 
further confirmed by the one-to-one comparison of the 3D vector magnetic field in the volume as defined 
per \eqss{Eq-Error-Bstrength}{Eq-Error-BCWL} and reported in \tab{Tab-BF-errors-LM}. Indeed, we find that 
the relative error for the magnetic field strength is below $0.06 \%$, hence very low, 
while the error on the magnetic 
field direction is below $0.05^{\circ}$ (whether analyzed from \eq{Eq-Error-Bangle} or \eq{Eq-Error-BCWL}).

%
%
The high accuracy on retrieving the ground-truth 3D magnetic field translates into a good accuracy for 
all magnetic field-related quantities considered in this paper. The errors on the electric current density 
strength and direction (\eqss{Eq-Error-Jstrength}{Eq-Error-JCWL}) are larger than for the magnetic field, 
as one would expect since the electric current density is related to the gradients of the magnetic field 
and spatial derivatives tend to enhance errors. However, the errors remain marginal since the electric 
current density strength is recovered with a relative error lower than $4 \%$ and its direction with an error 
lower than $1.4^{\circ}$. Similarly, we find good results for global quantities such as the free magnetic 
energy and the relative magnetic helicity, which are recovered with a relative error below $0.2 \%$. The same 
is true for the synthetic polarization data (\tab{Tab-BF-errors-LM}) and their related quantities (\eg $L/I$, 
$V/I$ and $AZ$ shown \fig{Fig-Stokes-comparison}), for which the errors are typically $10^{3}$ times smaller 
than the actual polarization signal.

%
%
The very high quality of these results is agnostic to the minimized MSE (\tab{Tab-BF-errors-LM}) and 
flux rope model used to produce the ground-truth synthetic observations (\tab{Tab-BF-errors-all-FRs}). 
We therefore conclude that the DOCFM approach allows us to accurately retrieve the 3D coronal magnetic 
field when the coronal polarization dataset to be fitted originates from a solution of the generative 
model (here, the flux rope insertion method). The DOCFM framework thus opens new perspectives for coupling 
photospheric magnetic field data to coronal polarimetry in 3D reconstruction of the solar coronal magnetic 
field.


\section{Discussion} \label{sec:S-Discussion}

%
%
As previously mentioned, DOCFM is a framework we propose to couple existing 3D magnetic field 
reconstructions with coronal polarimetric observations. It is a model-data fitting approach of the 3D 
reconstruction of the coronal magnetic field. It combines a parametrized generative magnetic field 
model, forward modeling techniques, and an optimization method for finding the magnetic field 
parameters that minimize the differences between the real data and synthetic observables predicted 
for the magnetic field model.

%
%
Although we applied and tested the DOCFM approach using the flux rope insertion method and coronal 
polarimetry, it should be emphasized that the proposed framework was developed in a way that is agnostic 
to the generative magnetic field model, the type of coronal data used to constrain the 3D magnetic field 
(off-limb and on-disk), 
and the optimization method (although ROAM was specifically designed for the DOCFM to be tractable 
even with computationally-heavy generative models having 3 parameters or more). The DOCFM is thus 
general enough to be used with any type of generative magnetic field model, whether it is, 
\eg an extrapolation method \citep[including flux rope insertion;][]{Wheatland00,vanBallegooijen04,Wiegelmann04,Valori05,Amari10,Contopoulos11,Titov14}, 
an MHD model \citep[\eg][]{Mikic99,Inoue11,Feng12,Zhu13}, or even an analytical one. 
The only constraint is that the generative model must be parametrized either through, \eg its magnetic 
field or its electric currents. Similarly, DOCFM can be used with any type of solar data, 
both off-limb and on-disk, provided 
that it can be forward modeled. This is the case for, \eg white-light/EUV/X-ray imaging or radio 
polarimetry through, \eg the FORWARD IDL suite \citep[see][and references therein]{Gibson16}.

%
%
In this paper, we showed that the DOCFM framework is applicable and allows one to 
accurately retrieve the 3D coronal magnetic field when the coronal synthetic observations 
are created from a known ground-truth physical state solution of the parametrized generative model. 
It should be emphasized, though, that DOCFM 
does not guarantee that the obtained solution will be the true, 3D, coronal magnetic field in both 
observational and -- more general -- modeling applications. Indeed, the ability of the DOCFM to provide 
the true coronal magnetic field entirely depends on the ability of the generative model to reproduce 
the true coronal magnetic field. That point will be addressed and further discussed in Paper II 
{(in preparation).} 
Thus, the DOCFM method is to be seen as a tool to directly include more observational constraints 
in extrapolations/reconstructions of the 3D coronal magnetic field, with the goal of obtaining 
better solutions. At this point, it should be mentioned that applying the DOCFM approach 
with un-parametrized extrapolation/reconstruction methods can be achieved without modifying 
the generative model technique, as long as parametrization is achieved through, \eg the photospheric 
vector magnetic field. The DOCFM approach therefore opens interesting new perspectives for investigating 
whether it may help different extrapolation methods to converge towards much more similar solutions, 
as well as for developing new reconstruction techniques to more accurately and robustly derive the 3D 
coronal magnetic field and its properties (\eg 3D distribution of coronal electric currents, 
magnetic topology, free magnetic energy, relative magnetic helicity) by combining photospheric and coronal data. 

%
%
In the investigations presented here, the orientation of the flux rope with regard 
to the LOS was not considered as a parameter to be fitted, although it might affect 
the fitting results. Current observational applications of the DOCFM approach with off-limb 
coronal data and a flux rope insertion method requires the use of on-disk data, \eg magnetogram 
and EUV emission, prior to (resp. after) the passage at the West (resp. East) limb to determine 
both the photospheric feet and insertion path of the flux rope. Then, the cube containing 
the 3D magnetic field model can either be rigidly rotated to the limb for fitting the off-limb 
coronal data or evolved by means of a photospheric flux transport model to take into account 
the evolution due to both differential rotation and granular motions 
\citep[see \eg][and references therein]{Yeates07}. In both cases, the orientation of the flux rope 
axis with regard to the LOS should thus be very well constrained. Note also that potential fitting 
issues due to uncertainties on the flux rope orientation with regard to the LOS would be completely 
removed in the case of fitting on-disk coronal data and/or using multi-viewpoint simultaneous 
observations taken 90 degrees apart (\ie, typically when the ESA/Solar Orbiter will be in quadrature 
with Earth, thus providing photospheric vector magnetograms and EUV flux rope data from atop, 
while Earth will be providing off-limb coronal observations of the flux rope from the side).

%
%
Finally, we recall that the forward modeling of coronal polarimetry (and most other coronal data) 
requires a 3D plasma density and temperature model \citep{Gibson16}, which is not provided by NLFFF 
extrapolation/reconstruction models such as the one used in this paper 
but could be provided by MHD models. Although using ratios of 
Stokes images (\eg $Q/I$ instead of $Q$) should help reducing the dependency of the Stokes-related 
images, and hence of the DOCFM solution, to the plasma density model, it may not completely suppress 
it. For this reason, future investigations with more complex and realistic plasma density and temperature 
models will be required to fully characterize the sensitivity of the DOCFM 3D magnetic field solution 
to the plasma model. Such investigations will be particularly useful to determine how complex and 
realistic the plasma model really needs to be to reliably apply the DOCFM framework in observational 
studies. On the other hand, plasma solutions derived from tomographic inversions 
\citep[\eg][]{Frazin00,Frazin05,Kramar09,Barbey11,Guennou12,Huang12} and/or line intensity ratios 
may provide a good solution to that problem in observational applications.


\section{Conclusions} \label{sec:S-Conclusions}

%
%
Deriving the 3D magnetic field in the volume of the solar atmosphere is critical 
for improving our understanding of solar activity and evolution 
\citep[see reviews by \eg][and references therein]{Forbes06}. 
Most present solar magnetic field data rely on the inversion of 2D photospheric and 
chromospheric polarization measurements \citep[\eg][]{Lagg17}. Such data 
only provide surface vector magnetic fields at the lowest layers of the solar atmosphere. 
3D models, \eg in the form of NLFFF and/or MHD solutions 
\citep[see review by \eg][and references therein]{Wiegelmann14}, 
coupled to photospheric and/or chromospheric surface vector magnetograms are then 
required to construct an approximate 3D magnetic field solution in the solar corona. 
Unfortunately, the inferred 3D solution and its properties can strongly depend 
on the chosen 3D reconstruction method \citep{DeRosa09,DeRosa15,Yeates18}. On the other 
hand, coronal polarization data could provide additional information to better constrain 
the 3D coronal magnetic field \citep[\eg][]{Casini99,Rachmeler13,Gibson17}. However, 
combining them with 3D reconstruction methods is not straightforward. In particular, 
such measurements cannot -- in general -- be inverted into 2D maps of vector magnetic 
fields that could be directly integrated into reconstruction methods (see \eg 
line-of-sight integration issue and limitations of stereoscopic measurements 
in \sect{S-Introduction}).

%
%
In this paper, we introduced the Data-Optimized Coronal Field Model (DOCFM), a new 
framework for coupling existing 3D magnetic field reconstruction techniques with 
coronal polarimetric data. The DOCFM is a model-data fitting approach of the 3D 
magnetic field reconstruction problem. It relies on the parametrization of 3D 
reconstruction methods through their electric currents (\eg surface parametrization 
at the photosphere or volume parametrization) and the fine-tuning of the electric 
current-related parameters such that the polarization signal predicted for the magnetic 
field model \citep[\ie through forward modeling;][]{Gibson16} matches the real 
polarization data.

%
%
By applying it with the flux rope insertion method of \cite{vanBallegooijen04} 
and IR coronal polarimetry in the Fe XIII lines as observed by the CoMP, 
we have demonstrated the applicability of 
the DOCFM methodology. We showed that coronal polarimetric 
data contain enough information to constrain the 3D coronal magnetic field 
solution when coupled with a parametrized 3D flux rope insertion method. 
While it does not guarantee that the inferred magnetic field solution is 
the true coronal magnetic field in observational applications, the DOCFM 
approach provides a means to force coronal magnetic field reconstructions 
to satisfy additional, common, coronal constraints. This framework therefore 
opens new perspectives for the exploitation of coronal polarimetry in 3D 
reconstructions of the solar coronal magnetic field.

%
%
Ideally, the application of the DOCFM methodology requires simultaneous 
measurements of photospheric and/or chromospheric vector magnetic fields 
viewed near solar disk center and off-limb coronal polarization. The Earth's vantage 
point alone does not permit co-spatial and co-temporal observations of this type. 
The upcoming ESA/Solar Orbiter space mission will enable such opportunities when 
Solar Orbiter will be in quadrature with Earth. Additionally, the DKIST will 
be sensitive enough to measure Zeeman-induced, coronal circular polarization 
at the limb \citep[\eg][]{Keil11}. Coordinated observations between Solar Orbiter 
and the ground-based CoMP 
and DKIST will then provide unique datasets to compare 3D magnetic field reconstruction 
methods and more reliably infer the 3D properties of the coronal magnetic field that 
trigger solar flares and CMEs.


\begin{acknowledgements}
{We thank the anonymous referee and Rebecca Centeno Elliott for a careful consideration 
of the manuscript and helpful comments.} 
K.D., A.S., S.G., Y.F. and E.D.L. acknowledge support from the Air Force Office 
of Scientific Research, FA9550-15-1-0030. This work was further supported 
by the Computational and Information Systems Laboratory, and the High Altitude 
Observatory. K.D. gratefully acknowledges the support of the french Centre National 
d'\'Etudes Spatiales. The calculations presented in this paper were performed using 
HPC resources from CALMIP (Grant 2018-2019 [P1504]). The National Center for Atmospheric 
Research is sponsored by the National Science Foundation.
\end{acknowledgements}

\bibliographystyle{aa}
      
\bibliography{DOCFM}  

\begin{thebibliography}{98}
\expandafter\ifx\csname natexlab\endcsname\relax\def\natexlab#1{#1}\fi

\bibitem[{{Alissandrakis}(1981)}]{Alissandrakis81}
{Alissandrakis}, C.~E. 1981, \aap, 100, 197

\bibitem[{{Amari} \& {Aly}(2010)}]{Amari10}
{Amari}, T. \& {Aly}, J.-J. 2010, \aap, 522, A52

\bibitem[{{Amari} {et~al.}(2013){Amari}, {Aly}, {Canou}, \& {Mikic}}]{Amari13}
{Amari}, T., {Aly}, J.-J., {Canou}, A., \& {Mikic}, Z. 2013, \aap, 553, A43

\bibitem[{{Amari} {et~al.}(2006){Amari}, {Boulmezaoud}, \& {Aly}}]{Amari06}
{Amari}, T., {Boulmezaoud}, T.~Z., \& {Aly}, J.~J. 2006, \aap, 446, 691

\bibitem[{{Amari} {et~al.}(2018){Amari}, {Canou}, {Aly}, {Delyon}, \&
  {Alauzet}}]{Amari18}
{Amari}, T., {Canou}, A., {Aly}, J.-J., {Delyon}, F., \& {Alauzet}, F. 2018,
  \nat, 554, 211

\bibitem[{{Antiochos} {et~al.}(1999){Antiochos}, {DeVore}, \&
  {Klimchuk}}]{Antiochos99}
{Antiochos}, S.~K., {DeVore}, C.~R., \& {Klimchuk}, J.~A. 1999, \apj, 510, 485

\bibitem[{{Arnaud} \& {Newkirk}(1987)}]{Arnaud87}
{Arnaud}, J. \& {Newkirk}, Jr., G. 1987, \aap, 178, 263

\bibitem[{{Aulanier} {et~al.}(2012){Aulanier}, {Janvier}, \&
  {Schmieder}}]{Aulanier12}
{Aulanier}, G., {Janvier}, M., \& {Schmieder}, B. 2012, \aap, 543, A110

\bibitem[{{Barbey} {et~al.}(2013){Barbey}, {Guennou}, \&
  {Auch{\`e}re}}]{Barbey11}
{Barbey}, N., {Guennou}, C., \& {Auch{\`e}re}, F. 2013, \solphys, 283, 227

\bibitem[{{Bateman}(1978)}]{Bateman78}
{Bateman}, G. 1978, {MHD instabilities}

\bibitem[{{B{\c a}k-St{\c e}{\'s}licka} {et~al.}(2013){B{\c a}k-St{\c
  e}{\'s}licka}, {Gibson}, {Fan}, {Bethge}, {Forland}, \&
  {Rachmeler}}]{BakSteslicka13}
{B{\c a}k-St{\c e}{\'s}licka}, U., {Gibson}, S.~E., {Fan}, Y., {et~al.} 2013,
  \apjl, 770, L28

\bibitem[{{Bobra} {et~al.}(2008){Bobra}, {van Ballegooijen}, \&
  {DeLuca}}]{Bobra08}
{Bobra}, M.~G., {van Ballegooijen}, A.~A., \& {DeLuca}, E.~E. 2008, \apj, 672,
  1209

\bibitem[{{Bommier} \& {Sahal-Brechot}(1982)}]{Bommier82}
{Bommier}, V. \& {Sahal-Brechot}, S. 1982, \solphys, 78, 157

\bibitem[{{Casini} \& {Judge}(1999)}]{Casini99}
{Casini}, R. \& {Judge}, P.~G. 1999, \apj, 522, 524

\bibitem[{{Centeno} {et~al.}(2010){Centeno}, {Trujillo Bueno}, \& {Asensio
  Ramos}}]{Centeno10}
{Centeno}, R., {Trujillo Bueno}, J., \& {Asensio Ramos}, A. 2010, \apj, 708,
  1579

\bibitem[{{Charvin}(1965)}]{Charvin65}
{Charvin}, P. 1965, Annales d'Astrophysique, 28, 877

\bibitem[{{Contopoulos} {et~al.}(2011){Contopoulos}, {Kalapotharakos}, \&
  {Georgoulis}}]{Contopoulos11}
{Contopoulos}, I., {Kalapotharakos}, C., \& {Georgoulis}, M.~K. 2011, \solphys,
  269, 351

\bibitem[{{Dalmasse} {et~al.}(2016){Dalmasse}, {Nychka}, {Gibson}, {Fan}, \&
  {Flyer}}]{Dalmasse16}
{Dalmasse}, K., {Nychka}, D., {Gibson}, S., {Fan}, Y., \& {Flyer}, N. 2016,
  Frontiers in Astronomy and Space Sciences, 3, 24

\bibitem[{{De Rosa} {et~al.}(2009){De Rosa}, {Schrijver}, {Barnes}, {Leka},
  {Lites}, {Aschwanden}, {Amari}, {Canou}, {McTiernan}, {R{\'e}gnier},
  {Thalmann}, {Valori}, {Wheatland}, {Wiegelmann}, {Cheung}, {Conlon},
  {Fuhrmann}, {Inhester}, \& {Tadesse}}]{DeRosa09}
{De Rosa}, M.~L., {Schrijver}, C.~J., {Barnes}, G., {et~al.} 2009, \apj, 696,
  1780

\bibitem[{{DeRosa} {et~al.}(2015){DeRosa}, {Wheatland}, {Leka}, {Barnes},
  {Amari}, {Canou}, {Gilchrist}, {Thalmann}, {Valori}, {Wiegelmann},
  {Schrijver}, {Malanushenko}, {Sun}, \& {R{\'e}gnier}}]{DeRosa15}
{DeRosa}, M.~L., {Wheatland}, M.~S., {Leka}, K.~D., {et~al.} 2015, \apj, 811,
  107

\bibitem[{{Fan}(2012)}]{Fan12}
{Fan}, Y. 2012, \apj, 758, 60

\bibitem[{{Feng} {et~al.}(2012){Feng}, {Jiang}, {Xiang}, {Zhao}, \&
  {Wu}}]{Feng12}
{Feng}, X., {Jiang}, C., {Xiang}, C., {Zhao}, X., \& {Wu}, S.~T. 2012, \apj,
  758, 62

\bibitem[{{Forbes}(2000)}]{Forbes00}
{Forbes}, T.~G. 2000, \jgr, 105, 23153

\bibitem[{{Forbes} {et~al.}(2006){Forbes}, {Linker}, {Chen}, {Cid}, {K{\'o}ta},
  {Lee}, {Mann}, {Miki{\'c}}, {Potgieter}, {Schmidt}, {Siscoe}, {Vainio},
  {Antiochos}, \& {Riley}}]{Forbes06}
{Forbes}, T.~G., {Linker}, J.~A., {Chen}, J., {et~al.} 2006, \ssr, 123, 251

\bibitem[{{Frazin}(2000)}]{Frazin00}
{Frazin}, R.~A. 2000, \apj, 530, 1026

\bibitem[{{Frazin} {et~al.}(2005){Frazin}, {Kamalabadi}, \& {Weber}}]{Frazin05}
{Frazin}, R.~A., {Kamalabadi}, F., \& {Weber}, M.~A. 2005, \apj, 628, 1070

\bibitem[{{Freeland} \& {Handy}(1998)}]{Freeland98}
{Freeland}, S.~L. \& {Handy}, B.~N. 1998, \solphys, 182, 497

\bibitem[{{Gary}(2001)}]{Gary01}
{Gary}, G.~A. 2001, \solphys, 203, 71

\bibitem[{{Gibson}(2015)}]{Gibson15}
{Gibson}, S. 2015, in Astrophysics and Space Science Library, Vol. 415, Solar
  Prominences, ed. J.-C. {Vial} \& O.~{Engvold}, 323

\bibitem[{{Gibson} {et~al.}(2016){Gibson}, {Kucera}, {White}, {Dove}, {Fan},
  {Forland}, {Rachmeler}, {Downs}, \& {Reeves}}]{Gibson16}
{Gibson}, S., {Kucera}, T., {White}, S., {et~al.} 2016, Frontiers in Astronomy
  and Space Sciences, 3, 8

\bibitem[{{Gibson} {et~al.}(2017){Gibson}, {Dalmasse}, {Rachmeler}, {De Rosa},
  {Tomczyk}, {de Toma}, {Burkepile}, \& {Galloy}}]{Gibson17}
{Gibson}, S.~E., {Dalmasse}, K., {Rachmeler}, L.~A., {et~al.} 2017, \apjl, 840,
  L13

\bibitem[{{Gorbachev} \& {Somov}(1989)}]{Gorbachev89}
{Gorbachev}, V.~S. \& {Somov}, B.~V. 1989, \sovast, 33, 57

\bibitem[{{Grad} \& {Rubin}(1958)}]{Grad58}
{Grad}, H. \& {Rubin}, H. 1958, in Proc. of the Second UN International Atomic
  Energy Conf. 31, Peaceful Uses of Atomic Energy

\bibitem[{{Guennou} {et~al.}(2012){Guennou}, {Auch{\`e}re}, {Soubri{\'e}},
  {Bocchialini}, {Parenti}, \& {Barbey}}]{Guennou12}
{Guennou}, C., {Auch{\`e}re}, F., {Soubri{\'e}}, E., {et~al.} 2012, \apjs, 203,
  25

\bibitem[{{Hanle}(1924)}]{Hanle24}
{Hanle}, W. 1924, Zeitschrift fur Physik, 30, 93

\bibitem[{{Harvey}(1969)}]{Harvey69}
{Harvey}, J.~W. 1969, PhD thesis

\bibitem[{{Hood} \& {Priest}(1981)}]{Hood81}
{Hood}, A.~W. \& {Priest}, E.~R. 1981, Geophysical and Astrophysical Fluid
  Dynamics, 17, 297

\bibitem[{{Huang} {et~al.}(2012){Huang}, {Frazin}, {Landi}, {Manchester},
  {V{\'a}squez}, \& {Gombosi}}]{Huang12}
{Huang}, Z., {Frazin}, R.~A., {Landi}, E., {et~al.} 2012, \apj, 755, 86

\bibitem[{{Iman} {et~al.}(1981){Iman}, {Helton}, \& {Campbell}}]{Iman81}
{Iman}, R.~L., {Helton}, J.~C., \& {Campbell}, J.~E. 1981, Journal of Quality
  Technology, 13, 174

\bibitem[{{Inoue} {et~al.}(2011){Inoue}, {Kusano}, {Magara}, {Shiota}, \&
  {Yamamoto}}]{Inoue11}
{Inoue}, S., {Kusano}, K., {Magara}, T., {Shiota}, D., \& {Yamamoto}, T.~T.
  2011, \apj, 738, 161

\bibitem[{{Inoue} {et~al.}(2012){Inoue}, {Magara}, {Watari}, \&
  {Choe}}]{Inoue12}
{Inoue}, S., {Magara}, T., {Watari}, S., \& {Choe}, G.~S. 2012, \apj, 747, 65

\bibitem[{{Jiang} {et~al.}(2016){Jiang}, {Wu}, {Yurchyshyn}, {Wang}, {Feng}, \&
  {Hu}}]{Jiang16}
{Jiang}, C., {Wu}, S.~T., {Yurchyshyn}, V., {et~al.} 2016, \apj, 828, 62

\bibitem[{{Judge}(1998)}]{Judge98}
{Judge}, P.~G. 1998, \apj, 500, 1009

\bibitem[{{Judge}(2007)}]{Judge07}
{Judge}, P.~G. 2007, \apj, 662, 677

\bibitem[{{Judge} {et~al.}(2006){Judge}, {Low}, \& {Casini}}]{Judge06}
{Judge}, P.~G., {Low}, B.~C., \& {Casini}, R. 2006, \apj, 651, 1229

\bibitem[{{Karna} {et~al.}(submitted){Karna}, {Savcheva}, {Dalmasse}, {Gibson},
  {Tassev}, {de Toma}, \& {DeLuca}}]{KarnaSub}
{Karna}, N., {Savcheva}, A., {Dalmasse}, K., {et~al.} submitted, \apj

\bibitem[{{Keil} {et~al.}(2011){Keil}, {Rimmele}, {Wagner}, {Elmore}, \& {ATST
  Team}}]{Keil11}
{Keil}, S.~L., {Rimmele}, T.~R., {Wagner}, J., {Elmore}, D., \& {ATST Team}.
  2011, in Astronomical Society of the Pacific Conference Series, Vol. 437,
  Solar Polarization 6, ed. J.~R. {Kuhn}, D.~M. {Harrington}, H.~{Lin}, S.~V.
  {Berdyugina}, J.~{Trujillo-Bueno}, S.~L. {Keil}, \& T.~{Rimmele}, 319

\bibitem[{{Kramar} {et~al.}(2013){Kramar}, {Inhester}, {Lin}, \&
  {Davila}}]{Kramar13}
{Kramar}, M., {Inhester}, B., {Lin}, H., \& {Davila}, J. 2013, \apj, 775, 25

\bibitem[{{Kramar} {et~al.}(2009){Kramar}, {Jones}, {Davila}, {Inhester}, \&
  {Mierla}}]{Kramar09}
{Kramar}, M., {Jones}, S., {Davila}, J., {Inhester}, B., \& {Mierla}, M. 2009,
  \solphys, 259, 109

\bibitem[{{Kramar} {et~al.}(2016){Kramar}, {Lin}, \& {Tomczyk}}]{Kramar16}
{Kramar}, M., {Lin}, H., \& {Tomczyk}, S. 2016, \apjl, 819, L36

\bibitem[{{Kuhn} {et~al.}(1996){Kuhn}, {Penn}, \& {Mann}}]{Kuhn96}
{Kuhn}, J.~R., {Penn}, M.~J., \& {Mann}, I. 1996, \apjl, 456, L67

\bibitem[{{Kusano} {et~al.}(2012){Kusano}, {Bamba}, {Yamamoto}, {Iida},
  {Toriumi}, \& {Asai}}]{Kusano12}
{Kusano}, K., {Bamba}, Y., {Yamamoto}, T.~T., {et~al.} 2012, \apj, 760, 31

\bibitem[{{Lagg} {et~al.}(2017){Lagg}, {Lites}, {Harvey}, {Gosain}, \&
  {Centeno}}]{Lagg17}
{Lagg}, A., {Lites}, B., {Harvey}, J., {Gosain}, S., \& {Centeno}, R. 2017,
  \ssr, 210, 37

\bibitem[{{Lin} {et~al.}(2004){Lin}, {Kuhn}, \& {Coulter}}]{Lin04}
{Lin}, H., {Kuhn}, J.~R., \& {Coulter}, R. 2004, \apjl, 613, L177

\bibitem[{{Lin} {et~al.}(2000){Lin}, {Penn}, \& {Tomczyk}}]{Lin00}
{Lin}, H., {Penn}, M.~J., \& {Tomczyk}, S. 2000, \apjl, 541, L83

\bibitem[{{Low}(1996)}]{Low96}
{Low}, B.~C. 1996, \solphys, 167, 217

\bibitem[{{Malanushenko} {et~al.}(2012){Malanushenko}, {Schrijver}, {DeRosa},
  {Wheatland}, \& {Gilchrist}}]{Malanushenko12}
{Malanushenko}, A., {Schrijver}, C.~J., {DeRosa}, M.~L., {Wheatland}, M.~S., \&
  {Gilchrist}, S.~A. 2012, \apj, 756, 153

\bibitem[{{McKay} {et~al.}(1979){McKay}, {Beckman}, \& {Conover}}]{McKay79}
{McKay}, M.~D., {Beckman}, R.~J., \& {Conover}, W.~J. 1979, Technometrics, 21,
  239

\bibitem[{{Miki{\'c}} {et~al.}(1999){Miki{\'c}}, {Linker}, {Schnack},
  {Lionello}, \& {Tarditi}}]{Mikic99}
{Miki{\'c}}, Z., {Linker}, J.~A., {Schnack}, D.~D., {Lionello}, R., \&
  {Tarditi}, A. 1999, Physics of Plasmas, 6, 2217

\bibitem[{{Morton} {et~al.}(2016){Morton}, {Tomczyk}, \& {Pinto}}]{Morton16}
{Morton}, R.~J., {Tomczyk}, S., \& {Pinto}, R.~F. 2016, \apj, 828, 89

\bibitem[{{Pariat} {et~al.}(2017){Pariat}, {Leake}, {Valori}, {Linton},
  {Zuccarello}, \& {Dalmasse}}]{Pariat17}
{Pariat}, E., {Leake}, J.~E., {Valori}, G., {et~al.} 2017, \aap, 601, A125

\bibitem[{{Penn}(2014)}]{Penn14}
{Penn}, M.~J. 2014, Living Reviews in Solar Physics, 11, 2

\bibitem[{{Plowman}(2014)}]{Plowman14}
{Plowman}, J. 2014, \apj, 792, 23

\bibitem[{{Priest}(2003)}]{Priest03}
{Priest}, E.~R. 2003, {Solar magnetohydrodynamics}, ed. B.~N. {Dwivedi} \&
  F.~b.~E.~N. {Parker}, 217--237

\bibitem[{{Querfeld}(1982)}]{Querfeld82}
{Querfeld}, C.~W. 1982, \apj, 255, 764

\bibitem[{{Rachmeler} {et~al.}(2013){Rachmeler}, {Gibson}, {Dove}, {DeVore}, \&
  {Fan}}]{Rachmeler13}
{Rachmeler}, L.~A., {Gibson}, S.~E., {Dove}, J.~B., {DeVore}, C.~R., \& {Fan},
  Y. 2013, \solphys, 288, 617

\bibitem[{{Rachmeler} {et~al.}(2014){Rachmeler}, {Platten}, {Bethge}, {Seaton},
  \& {Yeates}}]{Rachmeler14}
{Rachmeler}, L.~A., {Platten}, S.~J., {Bethge}, C., {Seaton}, D.~B., \&
  {Yeates}, A.~R. 2014, \apjl, 787, L3

\bibitem[{{Raouafi} {et~al.}(2016){Raouafi}, {Riley}, {Gibson}, {Fineschi}, \&
  {Solanki}}]{Raouafi16}
{Raouafi}, N.~E., {Riley}, P., {Gibson}, S., {Fineschi}, S., \& {Solanki},
  S.~K. 2016, Frontiers in Astronomy and Space Sciences, 3, 20

\bibitem[{{Sahal-Brechot} {et~al.}(1977){Sahal-Brechot}, {Bommier}, \&
  {Leroy}}]{SahalBrechot77}
{Sahal-Brechot}, S., {Bommier}, V., \& {Leroy}, J.~L. 1977, \aap, 59, 223

\bibitem[{{Savcheva} {et~al.}(2016){Savcheva}, {Pariat}, {McKillop},
  {McCauley}, {Hanson}, {Su}, \& {DeLuca}}]{Savcheva16}
{Savcheva}, A., {Pariat}, E., {McKillop}, S., {et~al.} 2016, \apj, 817, 43

\bibitem[{{Savcheva} {et~al.}(2012){Savcheva}, {van Ballegooijen}, \&
  {DeLuca}}]{Savcheva12}
{Savcheva}, A.~S., {van Ballegooijen}, A.~A., \& {DeLuca}, E.~E. 2012, \apj,
  744, 78

\bibitem[{{Schrijver} {et~al.}(2005){Schrijver}, {De Rosa}, {Title}, \&
  {Metcalf}}]{Schrijver05}
{Schrijver}, C.~J., {De Rosa}, M.~L., {Title}, A.~M., \& {Metcalf}, T.~R. 2005,
  \apj, 628, 501

\bibitem[{{Shibata} \& {Magara}(2011)}]{Shibata11}
{Shibata}, K. \& {Magara}, T. 2011, Living Reviews in Solar Physics, 8, 6

\bibitem[{{Somov} \& {Verneta}(1993)}]{Somov93}
{Somov}, B.~V. \& {Verneta}, A.~I. 1993, \ssr, 65, 253

\bibitem[{{Su} {et~al.}(2011){Su}, {Surges}, {van Ballegooijen}, {DeLuca}, \&
  {Golub}}]{Su11}
{Su}, Y., {Surges}, V., {van Ballegooijen}, A., {DeLuca}, E., \& {Golub}, L.
  2011, \apj, 734, 53

\bibitem[{{Titov} {et~al.}(2018){Titov}, {Downs}, {Miki{\'c}}, {T{\"o}r{\"o}k},
  {Linker}, \& {Caplan}}]{Titov18}
{Titov}, V.~S., {Downs}, C., {Miki{\'c}}, Z., {et~al.} 2018, \apjl, 852, L21

\bibitem[{{Titov} {et~al.}(2014){Titov}, {T{\"o}r{\"o}k}, {Mikic}, \&
  {Linker}}]{Titov14}
{Titov}, V.~S., {T{\"o}r{\"o}k}, T., {Mikic}, Z., \& {Linker}, J.~A. 2014,
  \apj, 790, 163

\bibitem[{{Tomczyk} {et~al.}(2008){Tomczyk}, {Card}, {Darnell}, {Elmore},
  {Lull}, {Nelson}, {Streander}, {Burkepile}, {Casini}, \& {Judge}}]{Tomczyk08}
{Tomczyk}, S., {Card}, G.~L., {Darnell}, T., {et~al.} 2008, \solphys, 247, 411

\bibitem[{{Tomczyk} {et~al.}(2016){Tomczyk}, {Landi}, {Burkepile}, {Casini},
  {DeLuca}, {Fan}, {Gibson}, {Lin}, {McIntosh}, {Solomon}, {Toma}, {Wijn}, \&
  {Zhang}}]{Tomczyk16}
{Tomczyk}, S., {Landi}, E., {Burkepile}, J.~T., {et~al.} 2016, Journal of
  Geophysical Research (Space Physics), 121, 7470

\bibitem[{{Tziotziou} {et~al.}(2012){Tziotziou}, {Georgoulis}, \&
  {Raouafi}}]{Tziotziou12}
{Tziotziou}, K., {Georgoulis}, M.~K., \& {Raouafi}, N.-E. 2012, \apjl, 759, L4

\bibitem[{{Valori} {et~al.}(2005){Valori}, {Kliem}, \& {Keppens}}]{Valori05}
{Valori}, G., {Kliem}, B., \& {Keppens}, R. 2005, \aap, 433, 335

\bibitem[{{Valori} {et~al.}(2010){Valori}, {Kliem}, {T{\"o}r{\"o}k}, \&
  {Titov}}]{Valori10}
{Valori}, G., {Kliem}, B., {T{\"o}r{\"o}k}, T., \& {Titov}, V.~S. 2010, \aap,
  519, A44

\bibitem[{{van Ballegooijen}(2004)}]{vanBallegooijen04}
{van Ballegooijen}, A.~A. 2004, \apj, 612, 519

\bibitem[{{van Ballegooijen} {et~al.}(2000){van Ballegooijen}, {Priest}, \&
  {Mackay}}]{vanBallegooijen00}
{van Ballegooijen}, A.~A., {Priest}, E.~R., \& {Mackay}, D.~H. 2000, \apj, 539,
  983

\bibitem[{{Van Vleck}(1925)}]{VanVleck25}
{Van Vleck}, J.~H. 1925, Proceedings of the National Academy of Science, 11,
  612

\bibitem[{{Wheatland}(2007)}]{Wheatland07}
{Wheatland}, M.~S. 2007, \solphys, 245, 251

\bibitem[{{Wheatland} {et~al.}(2000){Wheatland}, {Sturrock}, \&
  {Roumeliotis}}]{Wheatland00}
{Wheatland}, M.~S., {Sturrock}, P.~A., \& {Roumeliotis}, G. 2000, \apj, 540,
  1150

\bibitem[{{Wiegelmann}(2004)}]{Wiegelmann04}
{Wiegelmann}, T. 2004, \solphys, 219, 87

\bibitem[{{Wiegelmann} \& {Inhester}(2010)}]{Wiegelmann10}
{Wiegelmann}, T. \& {Inhester}, B. 2010, \aap, 516, A107

\bibitem[{{Wiegelmann} \& {Sakurai}(2012)}]{Wiegelmann12R}
{Wiegelmann}, T. \& {Sakurai}, T. 2012, Living Reviews in Solar Physics, 9, 5

\bibitem[{{Wiegelmann} {et~al.}(2012){Wiegelmann}, {Thalmann}, {Inhester},
  {Tadesse}, {Sun}, \& {Hoeksema}}]{Wiegelmann12}
{Wiegelmann}, T., {Thalmann}, J.~K., {Inhester}, B., {et~al.} 2012, \solphys,
  281, 37

\bibitem[{{Wiegelmann} {et~al.}(2014){Wiegelmann}, {Thalmann}, \&
  {Solanki}}]{Wiegelmann14}
{Wiegelmann}, T., {Thalmann}, J.~K., \& {Solanki}, S.~K. 2014, \aapr, 22, 78

\bibitem[{{Yang} {et~al.}(1986){Yang}, {Sturrock}, \& {Antiochos}}]{Yang86}
{Yang}, W.~H., {Sturrock}, P.~A., \& {Antiochos}, S.~K. 1986, \apj, 309, 383

\bibitem[{{Yeates}(2014)}]{Yeates14}
{Yeates}, A.~R. 2014, \solphys, 289, 631

\bibitem[{{Yeates} {et~al.}(2018){Yeates}, {Amari}, {Contopoulos}, {Feng},
  {Mackay}, {Miki{\'c}}, {Wiegelmann}, {Hutton}, {Lowder}, {Morgan}, {Petrie},
  {Rachmeler}, {Upton}, {Canou}, {Chopin}, {Downs}, {Druckm{\"u}ller},
  {Linker}, {Seaton}, \& {T{\"o}r{\"o}k}}]{Yeates18}
{Yeates}, A.~R., {Amari}, T., {Contopoulos}, I., {et~al.} 2018, \ssr, 214, 99

\bibitem[{{Yeates} {et~al.}(2007){Yeates}, {Mackay}, \& {van
  Ballegooijen}}]{Yeates07}
{Yeates}, A.~R., {Mackay}, D.~H., \& {van Ballegooijen}, A.~A. 2007, \solphys,
  245, 87

\bibitem[{{Zhu} {et~al.}(2013){Zhu}, {Wang}, {Du}, \& {Fan}}]{Zhu13}
{Zhu}, X.~S., {Wang}, H.~N., {Du}, Z.~L., \& {Fan}, Y.~L. 2013, \apj, 768, 119

\bibitem[{{Zuccarello} {et~al.}(2018){Zuccarello}, {Pariat}, {Valori}, \&
  {Linan}}]{Zuccarello18}
{Zuccarello}, F.~P., {Pariat}, E., {Valori}, G., \& {Linan}, L. 2018, \apj,
  863, 41

\end{thebibliography}

\IfFileExists{\jobname.bbl}{} {\typeout{}
\typeout{****************************************************}
\typeout{****************************************************}
\typeout{** Please run "bibtex \jobname" to obtain} \typeout{**
the bibliography and then re-run LaTeX} \typeout{** twice to fix
the references !}
\typeout{****************************************************}
\typeout{****************************************************}
\typeout{}}


\end{document}